\documentclass[aps,a4paper,superscriptaddress,nofootinbib]{revtex4-1}
\usepackage{amsmath,amssymb,graphicx,multirow,dcolumn,bm,latexsym,soul,nicefrac}
\usepackage{epstopdf}
\usepackage{mathrsfs}
\usepackage{acronym}
\usepackage[colorlinks,linkcolor=blue,citecolor=blue,urlcolor=blue ]{hyperref}
\usepackage{ulem}
\usepackage{subcaption}
\usepackage[labelformat=simple,labelsep=period,skip=3pt]{caption}
\newcommand{\nn}{\nonumber}
\newcommand{\GABE}{{GABE}}

\newcommand{\Msun}{\,{\rm M}_\odot}
\newcommand{\hMpc}{\,h^{-1}\,{\rm Mpc}}

\newcommand{\kmsMpc}{\,{\rm km}\,\,{\rm s}^{-1}\,{\rm Mpc^{-1}}}
\newcommand{\Sa}{\,{\rm m}\,{\rm s}^{-2}\,{\rm Hz}^{-1/2}}
\newcommand{\Sx}{\,{\rm m}\,{\rm Hz}^{-1/2}}

\newcommand{\TSS}{TianQin Research Center for Gravitational Physics \& School of Physics and Astronomy, Sun Yat-sen University, 2 Daxue Rd., Zhuhai 519082, China}
\newcommand{\HUST}{MOE Key Laboratory of Fundamental Physical Quantities Measurements, 
Hubei Key Laboratory of Gravitation and Quantum Physics, PGMF and School of Physics, 
Huazhong University of Science and Technology, Wuhan 430074, China}
\def\apj{Astrophysical Journal}                % Astrophysical Journal
\def\apjl{Astrophysical Journal}             % Astrophysical Journal, Letters

\def\mnras{MNRAS}            % Monthly Notices of the RAS
\def\prd{Phys.~Rev.~D}       % Physical Review D
\def\araa{ARA\&A}             % Annual Review of Astron and Astrophys
\def\nat{Nature}              % Nature
\def\aapr{A\&A~Rev.~}    % Astronomy and Astrophysics Reviews
\begin{document}

\title{Science with the TianQin observatory: Preliminary results on massive black hole binaries}

\author{Hai-Tian Wang}
\affiliation{\TSS}
%\affiliation{\SPA}
\author{Zhen Jiang}
\affiliation{Key Laboratory for Computational Astrophysics, National Astronomical Observatories, Chinese Academy of Sciences, Beijing, 100012, China}
\affiliation{School of Astronomy and Space Science, University of Chinese Academy of Sciences, Beijing 10039, China}
\author{Alberto Sesana}
\affiliation{School of Physics and Astronomy and Institute of Gravitational Wave Astronomy, University of Birmingham, Edgbaston, Birmingham B15 2TT, United Kingdom}
\author{Enrico Barausse}
\affiliation{Institut d'Astrophysique de Paris, CNRS \& Sorbonne
 Universit\'es, UMR 7095, 98 bis bd Arago, 75014 Paris, France}
 \affiliation{SISSA, Via Bonomea 265, 34136 Trieste, Italy and INFN Sezione di Trieste}
 \affiliation{IFPU - Institute for Fundamental Physics of the Universe, Via Beirut 2, 34014 Trieste, Italy}
\author{Shun-Jia Huang}
\affiliation{\TSS}
%\affiliation{\SPA}
\author{Yi-Fan Wang}
\affiliation{Department of Physics, The Chinese University of Hong Kong, Shatin, N.T., Hong Kong}
\author{Wen-Fan Feng}
\author{Yan Wang}
\affiliation{\HUST}
\author{Yi-Ming Hu}
\email{huyiming@sysu.edu.cn}
\affiliation{\TSS}
\author{Jianwei Mei}
\author{Jun Luo}
\affiliation{\TSS}
\affiliation{\HUST}
%\affiliation{\SPA}

\date{\today}

\begin{abstract}
We investigate the prospects of detecting gravitational waves from coalescing massive black hole binaries in the Universe with the TianQin observatory, a space-based gravitational wave interferometer proposed to be launched in the 2030s.
To frame the scientific scope of the mission, in this paper, we carry out a preliminary estimation of the signal-to-noise ratio, detection rate, and parameter estimation precision of massive black hole binaries detectable by TianQin.
In order to make our results as robust as possible, we consider several models of the growth history of massive black holes, exploring the effect of some key astrophysical prescriptions as well as the impact of the employed computational methods.
In the most optimistic model, TianQin can detect as many as approximately $60$ mergers per year.
If TianQin detects a merger at redshift of $15$, it will be capable of estimating its luminosity distance to within an accuracy of $10\%$; for a nearby event at redshift approximately $2$, TianQin can issue early warnings $24$ hours before coalescence, with a timing accuracy of around three hours and a sky localization ability of approximately $80$ deg$^2$, thus enabling multimessenger observations.
\end{abstract}

%\keywords{semi-analytic, massive black hole, merger rate, TianQin}

%\pacs{04.25.dg, 04.40.Nr, 04.70.-s, 04.70.Bw}

%%%%%%%%%%%%%%%%%%%%%%%%%%%%%%%%%%%%%%%%%%%%%%%%%%%%%%%%%%%%%%%%
\maketitle

\acrodef{GW}{gravitational wave}
\acrodef{LIGO}{Laser Interferometer Gravitational-Wave Observatory}
\acrodef{BNS}{binary neutron star}
\acrodef{TQ}{TianQin}
\acrodef{PTA}{Pulsar timing array}
\acrodef{MBH}{massive black hole}
\acrodef{GR}{general relativity}
\acrodef{PN}{post-Newtonian}
\acrodef{NR}{numerical relativity}
\acrodef{SNR}{signal-to-noise ratio}
\acrodef{PSD}{power spectrum density}
\acrodef{RSS}{root sum square}
\acrodef{FIM}{Fisher information matrix}
\acrodef{SPA}{stationary phase approximation}

\section{Introduction}\label{sec:intro}

% ground facility introduction
The field of \ac{GW} astrophysics has witnessed a series of breakthroughs in the past few years.
The historic first direct detection of a \ac{GW} signal was made by the two \ac{LIGO} detectors \cite{aLIGO} in Hanford, Washington, and Livingston, Louisiana \citep{PhysRevLett.116.061102}, followed by two other black hole mergers \citep{PhysRevLett.116.241103,PhysRevLett.118.221101}.
In the second observation run, more systems have been observed, for a total of 11 detections claimed to date \citep{LIGOScientific:2018mvr}.
A few of them have been captured by the full ground-based \ac{GW} detector network \citep{PhysRevLett.119.141101}, which includes the Advanced Virgo interferometer \cite{TheVirgo:2014hva}, dramatically increasing the sky location accuracy.
Most notably, on August 17, 2017, the detection of a binary neutron star merger \citep{2017PhRvL.119p1101A}, followed by a distinctive counterpart in the electromagnetic spectrum \citep{2017ApJ...848L..12A}, ushered in the era of multimessenger \ac{GW} astronomy.
This series of discoveries has greatly deepened our understanding of the Universe; we tested the nature of gravity in a new laboratory, we understand the properties of the stellar-mass compact objects at an unprecedented level, we revealed the origin of most heavy elements nucleosynthesis, we learned the expansion of the Universe through an independent method, and so on (see, e.g., Refs. \cite{2013LRR....16....7G,2013LRR....16....9Y,2015CQGra..32x3001B,Kalogera3017,2017ApJ...848L..17C,2018arXiv180605195B}).

However, at frequencies lower than approximately $10$ Hz, the \ac{GW} spectrum remains unexplored.
Pulsar timing arrays \cite{2015ApJ...813...65T,2016MNRAS.455.1751R,2016MNRAS.458.3341D,2016IPTA} are hunting the heaviest black hole binaries in the Universe at nanohertz frequencies, still without success, despite the steady progresses made in the last decade \cite{2015Sci...349.1522S,2015MNRAS.453.2576L,2018ApJ...859...47A}.
However, it is in the millihertz to hertz band that the greatest richness and diversity of \ac{GW} sources are expected \cite{2013GWN.....6....4A,Hu17}.
Among such sources, mergers of \ac{MBH} binaries with masses
between approximately $10^4 M_\odot$ and approximately $10^7 M_\odot$ are expected to be the loudest \citep{Klein16, 2015JPhCS.610a2001B, Hu17}.
Indeed, electromagnetic observations have revealed the ubiquitous existence of \acp{MBH} in the center of galaxies \cite{1995ARA&A..33..581K} and most notably the approximately $4\times 10^6 M_\odot$ black hole Sagittarius A* within our Milky Way, e.g, Ref. \cite{2018A&A...615L..15G}.
Within the hierarchical structure formation process predicted by the standard cosmology, also known as the Lambda cold dark matter ($\Lambda$CDM) model, whenever galaxies merge to form larger ones, the \acp{MBH} hosted at their centers sink toward the center of the newly formed system, eventually binding into a binary.
The black hole binary gradually loses orbital energy and angular momentum by interacting with the stars and (if present) with the gas in its vicinity, eventually reaching a phase dominated 
by \ac{GW} emission, which climaxes in the final coalescence \cite{1980Natur.287..307B}.

The millihertz to hertz band will be probed by space-based \ac{GW} observatories, of which the best studied case is the European Space Agency (ESA) led by Laser Interferometer Space Antenna (LISA) \cite{2017arXiv170200786A}, scheduled for launch in 2034. In this paper, we focus on the TianQin project \cite{Luo:2015ght}, which was first put forward in 2014. 
TianQin is a space-based \ac{GW} observatory with the goal of being launched in the 2030s. 
In its simplest form, TianQin will be a constellation of three satellites, on a common geocentric orbit with a radius of about $10^5$ km.
The three satellites are spaced evenly on the orbit to form a nearly normal triangle.
There are test masses in each satellite, and the satellites are drag-free controlled to suppress nongravitational disturbances, so that the test masses can follow geodesic motion as closely as possible.
Laser interferometry is utilized to measure the variation in the light path between pairs of test masses caused by the passing \acp{GW}.

In this paper, we investigate the prospects of detecting \acp{GW} from \ac{MBH} binaries with TianQin.
To make the result as robust as possible, we consider five different models for the growth history of \acp{MBH}, including two light-seed models and three heavy-seed models.
Among those, one light-seed model and one heavy-seed model are based on the Millennium-I cosmological simulation \cite{Springel05}, while the other three are based on the extended Press and Schechter (EPS) formalism \cite{1974ApJ...187..425P,Parkinson2008}. 
(Note, however, that the EPS formalism we employ was calibrated to reproduce the results of the Millennium-I simulation, cf. Ref. \cite{Parkinson2008}.)

We focus on the detection ability and parameter measurement accuracy of TianQin on the sources of \ac{MBH} binaries.
The scope of applying such detections for further research are left for separate studies like Ref. \cite{Shi2019}.

The paper is organized as follows.
In Sec. \ref{sec:model}, we introduce the models used to reconstruct the growth history of MBHs.
In Sec. \ref{sec:method}, we describe the \ac{GW} signal, the TianQin sensitivity, and the mathematical method employed to compute \acp{SNR} and parameter estimation precision.
In Sec. \ref{sec:result}, we present our main results, and a brief summary and outlook are provided in Sec. \ref{sec:summary}.
Throughout the paper, unless otherwise specified, we adopt geometric units $G = c = 1$.

%%%%%%%%%%%%%%%%%%%%%%%%%%%%%%%%%%%%%%%%%%%%%%%%%%%%%%%%%%%%%%%%%%%
\section{Models for massive black hole binaries}\label{sec:model}

There is consensus through observation that the majority of galaxies have MBHs at their centers, and it is believed that MBHs are deeply intertwined with their host galaxies \cite{Antonucci93,Urry95,2000MNRAS.311..576K,2005Natur.433..604D,Croton06,Yang:2018ycs}. 
In the hierarchical clustering scenario of structure formation, the merger of galaxies that gives rise to the cosmic structure  \cite{Davis85,White78} is intimately related to the merger history of the \acp{MBH} hosted at their centers \cite{1980Natur.287..307B}. 
With traditional observation methods, however, it is hard to study the innermost properties of \acp{MBH} (cf. Ref. \cite{Icecube18}) and to capture the merger of \ac{MBH} binaries on action is beyond our current capability. 
Hence, the observation of the \ac{GW} signals coming from \ac{MBH} binary systems can greatly help to shape our understanding of the Universe at all scales, from the phenomena happening in the strong dynamical field of a merging black hole pair to the physics that drive the merger of galaxies and their growth along the cosmic history. 

In the first \ac{MBH} growth scenario, the cooling of metal-free atomic and molecular gas leads to their falling into the gravitational potential well of the first dark matter halos, which results in the generation of heavy (i.e., greater than $100 M_\odot$) stars, known as the population III (popIII) stars.
Pop III stars eventually collapse into seed black holes with mass approximately $100 M_\odot$ at redshift about $15-20$ \cite{2001ApJ...551L..27M}. 
These seed black holes then started to grow through accretion, and evolving together with their host halos and galaxies, they eventually grew into the \acp{MBH} identified in the center of galaxies in the local Universe \citep{2002ApJ...564...23B,2002Sci...295...93A,2016MNRAS.458.3047P}. 
This scenario is identified as the light-seed model (L-seed) model. 

The L-seed model encounters difficulty in explaining the luminous quasars, usually attributed to \acp{MBH}, observed at redshifts as high as $z=7.5$, when the Universe was less than 1 Gyr old \cite{2018Natur.553..473B}. 
This problem is alleviated if the initial mass of the black hole seed is bigger, as is the case for the so-called heavy-seed models (H-seed) models. 
The main idea is that dissociation of molecular gas due to strong UV background suppresses the molecular cooling, leading to popIII star formation. 
In these conditions, only atomic cooling, which occurs at much higher temperatures and requires much larger gas masses, is effective. 
If the required conditions are met, an \ac{MBH} of about $10^4\sim10^6 M_\odot$ can be directly formed through the collapse of atomic gas in massive protogalactic disks at the very early Universe (redshift greater than approximately $10$) \citep{2003ApJ...596...34B,2006ApJ...650..669V,Begelman:2006db,Lodato:2006hw,2017NatAs...1E..75R}. 
Notice that there are also different flavors in this scenario (see reviews in Refs. \cite{2010A&ARv..18..279V,2018arXiv181012310W}). 

It is currently impossible to pinpoint the actual growth path of seed black holes via traditional electromagnetic observations because \acp{MBH} rapidly lose memory of their initial conditions by accreting ambient gas. 
However, since the L-seed model predicts mergers of \ac{MBH} binaries of masses in the range $10^2-10^4 M_\odot$, while the H-seed model does not, one can in principle use \ac{GW} observations to distinguish among the two seeding scenarios, especially through accurate determination of masses and distances\cite{2011PhRvD..83d4036S}.

The starting point of our work is to predict the population distribution of \ac{MBH} binaries based on astronomical and cosmological observations. 
Such a prediction typically involves three ingredients: i) a dark matter halo merger tree, ii) a galaxy formation and evolution model, and iii) a black hole accretion model. 
We shall consider two sets of models, which we call the Millennium-I-based models and the EPS-based models.

\subsection{Millennium-I-based models}
\label{Sec:millennium}

For the Millennium-I-based models \cite{Jiang2018}, the dark matter halo merger trees are derived from the Millennium-I cosmological simulation \cite{Springel05}. 
The Millennium-I simulation, being one of the most widely used N-body dark matter halo simulations, has a good balance between box size ($500 \hMpc$) and mass resolution ($8.6 \times 10^8 h^{-1}M_{\odot}$). 
It employs the WMAP1 cosmology, $\Omega_{\rm m} = 0.25$, $\Omega_{\rm b} = 0.045$, $\Omega_{\rm \Lambda} = 0.75$, $n = 1$, $\sigma_{\rm 8} = 0.9$, and $H_{\rm 0} = 73 \kmsMpc$, derived from a combined analysis of the 2dFGalaxy Redshift Survey \cite{Colless01} and the first-year WMAP data \cite{Spergel03}.
Using the Millennium-I simulation has the advantage of deriving the dark matter halo evolution from the first principles, adopting as few assumptions as possible.
The drawback is also obvious: the limitation from computation ability constrains its mass resolution.

The galaxy formation and evolution are assigned by the the semi-analytic model Galaxy Assembly with Binary Evolution ({\GABE}). 
The model consists of a set of analytical prescriptions, either empirical or based on simple physical assumptions, each of which represents a physical process in galaxy formation and evolution. 
All prescriptions are employed simultaneously into the backbone of the cold dark matter halos, and the result is calibrated to match a set of observations, including the stellar-mass function and black hole-bulge mass relationship. 
Semianalytic models of this type are efficient, self-consistent, and consistent with observations, making them powerful tools for exploring different routes of galaxy and \acp{MBH} formation and evolution. 
{\GABE} is a newly developed semi-analytic model, containing a full set of galactic physics recipes, including, more importantly for this study, reionisation, hot gas cooling, star formation, supernova feedback, black hole growth, active galactic nucleus (AGN) feedback, bar formation, tidal stripping, and dynamical friction. 
{\GABE} can accurately recover galaxy properties at low redshift. 
Apart from galaxy formation and evolution, {\GABE} also includes black hole accretion as we describe below. 
More details about {\GABE} will be available in Ref. \cite{Jiang2018}.

In the L-seed model we assume that seed black holes are stellar remnants of $M_{seed} \gtrsim 250 M_{\odot}$ \cite{2001ApJ...551L..27M} of popIII stars, and that they form in minihalos with $\mathrm{T_{vir} < 10^4 K}$. 
In order to match observations, the light seeds must have experienced episodic super-Eddington accretion \cite{2014ApJ...784L..38M,2015ApJ...804..148V}. 
In the H-seed model, baryonic matter forms protogalactic disks in halos with $\mathrm{T_{vir}>10^4 K}$ at $z \gtrsim 10$. 
UV photodissociation of molecular hydrogen prevents them from producing popIII, and they directly collapse into seed black hole with mass $10^5-10^6\Msun$ \citep{2003ApJ...596...34B,2006ApJ...650..669V,2017NatAs...1E..75R}.

The MBH seeding in {\GABE} is performed in a simple way: whenever a dark matter halo's virial mass exceeds a critical mass $M_{\rm vir,crit}$, a black hole with mass $M_{\rm seed}$ will be assigned from a log-normal distribution to the central galaxy in this halo. 
For the L-seed model, $M_{\rm vir,crit}=10^{7}\Msun$, and a log-mean of $\mu_{\rm seed}=10^2\Msun$; for the H-seed model, $M_{\rm vir,crit}=10^{10}\Msun$, and $\mu_{\rm seed} =10^5\Msun$. 
Note that both masses are below the halo mass resolution in Millennium-I simulation, which is $2.4 \times 10^{10} \Msun$, i.e., 20 times its particle mass. 
In practice, we then assign a seed with mass $M_{\rm seed}$ when a halo first appears in the simulation. 

The black hole growth in {\GABE} follows the models of Ref. \cite{Croton06}. 
The growth is divided into two channels: quasar mode and radio mode.
The quasar mode is the rapid growth of MBHs in the process of galaxy merger, during which the gravitational and gaseous environment is highly disturbed. 
Gas clouds fall into the central black hole, making super-Eddington accretion possible, causing the rapid growth of seeds into \acp{MBH}. 
Radio mode is the quiet accretion of a MBH, and the accretion rate is set to the Bondi rate \cite{Bondi52}. 
The Bondi accretion rate is proportional to the square of the black hole mass; thus, this radio mode is more efficient for more massive objects and can create radio lobes in $L*$ or more massive galaxies, such as the Milky Way. 
Reference \cite{Croton06} proposed an empirical formula of such radio mode accretion and proved that the radio feedback is not sensitive to the details of accretion models. 

In {\GABE}, \ac{MBH} mergers are triggered by galaxy mergers, and no time delay between these two mergers was considered. 
This is a simplification, as the two \acp{MBH} need time to sink to the the center of the merger remnant and to then dissipate their orbital energy and angular momentum before the final coalescence. 
Proceeding from larger to smaller separation, different physical mechanisms play leading roles, including dynamical friction within the stellar and gaseous background, three-body interactions and slingshot of stars intersecting the \acp{MBH} binary orbit, gravitational and viscous torques exerted by a putative massive circumbinary disk, and, finally, GW emission when the separation between the two \acp{MBH} gets down to megaparsec scales. 
These could cause time delays of a few billion years, e.g. Refs. \cite{2002MNRAS.331..935Y,2015ApJ...810...49V,2015MNRAS.454L..66S}. 
In order to quantify all these physical processes, physical information of the inner part of host galaxies is needed, such as central gas and stellar densities. 
However, the simple disk and bulge models employed in {\GABE} do not yet provide such structural information. 
Thus, time delay is not implemented in {\GABE} as of now.

\begin{figure}[htb]
\centering
\begin{subfigure}[b]{.5\linewidth}
\centering
\includegraphics[width=\textwidth,height=7cm]{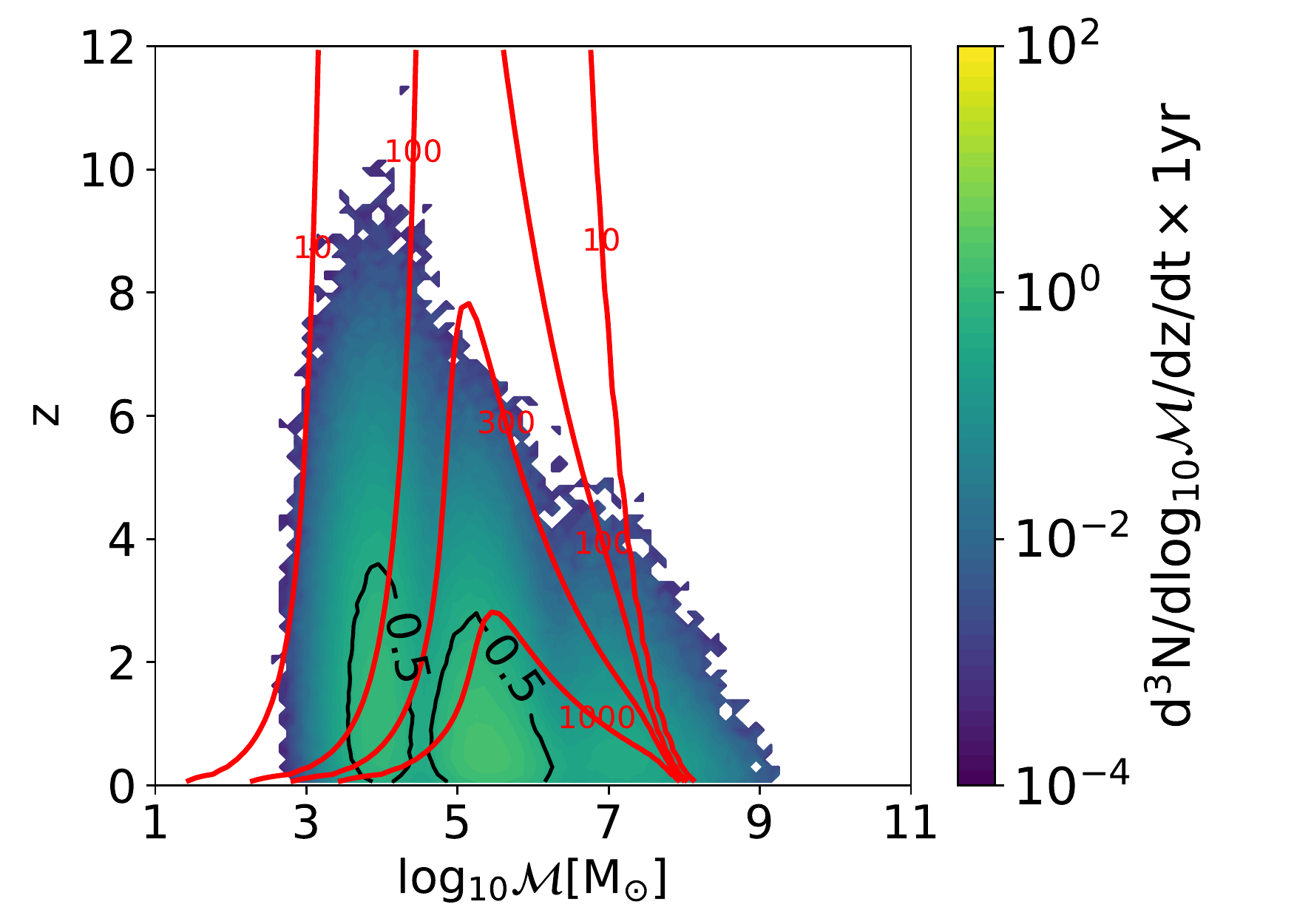}
\caption{H-seed}\label{fig_cm_red_SNR1}
\end{subfigure}%
\begin{subfigure}[b]{.5\linewidth}
\centering
\includegraphics[width=\textwidth,height=7cm]{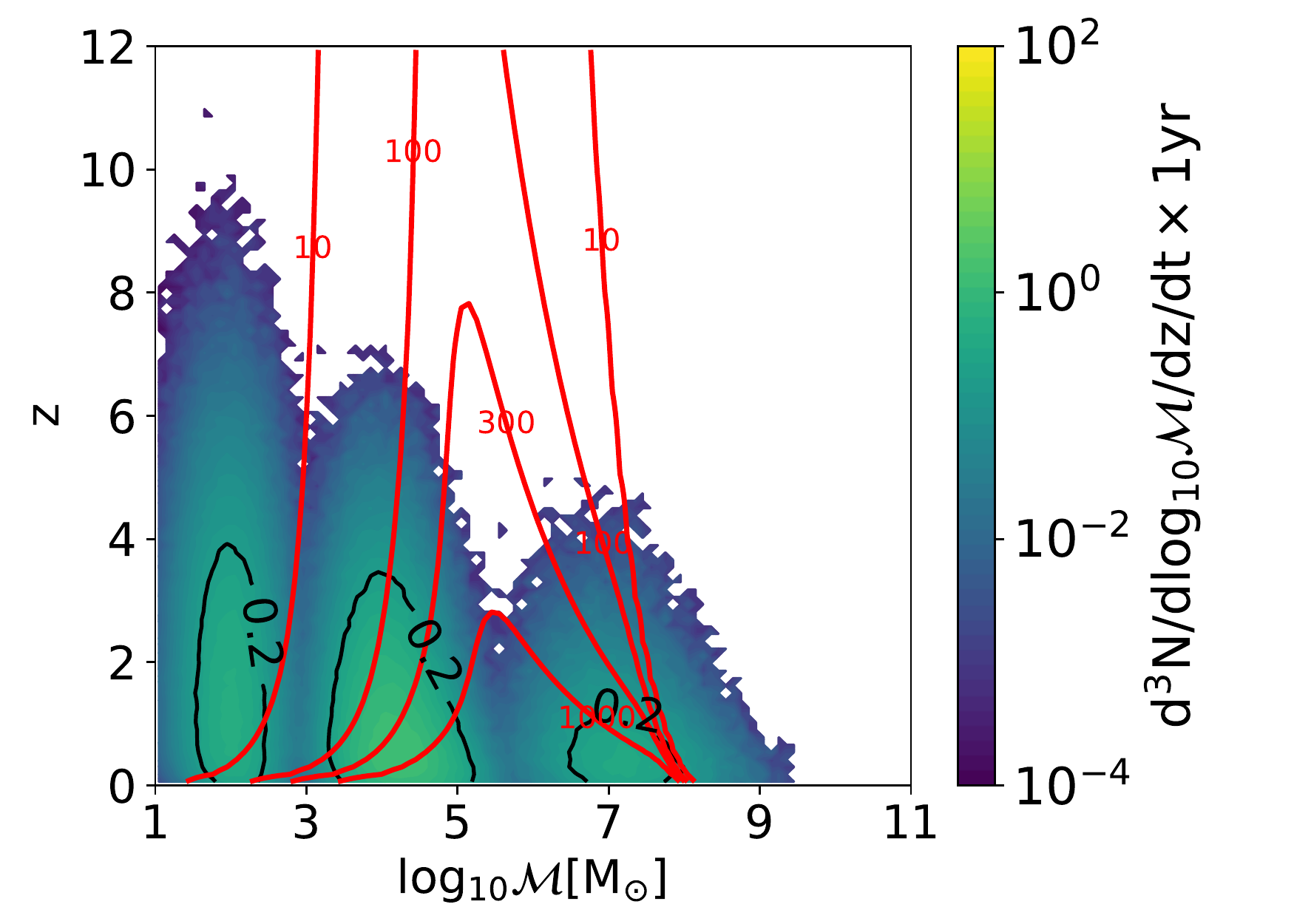}
\caption{L-seed}\label{fig_cm_red_SNR2}
\end{subfigure}%
\caption{The distribution of \acp{MBH} mergers over chirp mass and redshift from the heavy-seed model (left panel) and the light-seed model (right panel) in Ref. \cite{Jiang2018}.
  The red contour line represents the average SNR in the TianQin detector, assuming equal-mass binaries (details of calculation are discussed in Sec. \ref{sec:SNR}). The black lines represents contours on the differential number of mergers throughout the cosmic history.}
\label{fig_cm_red_SNR}
\end{figure}

The differential number of mergers throughout the cosmos over chirp mass $\mathcal{M}=(m_1m_2)^{3/5}/(m_1+m_2)^{1/5}$ ($m_{1,2}$ are component masses) and redshift of merging binaries for these two models are shown in Fig. \ref{fig_cm_red_SNR} together with \ac{SNR}, the definition of which will be discussed in detail in Sec. \ref{sec:SNR}.
For the H-seed model, we can see that the black hole mergers start to appear at $z\sim12$ with chirp mass of approximately $10^{5} \Msun$, which is the adopted seed mass. 
Toward lower redshift, along with the hierarchical evolution of the Universe, more and more MBH mergers appear, and their chirp mass increases. 
The plot reveals three chirp mass subpopulations. 
For the L-seed model, the distribution has the same trend, but with a lower chirp mass limit, and there is a larger gap between different chirp mass populations. 
These three chirp mass populations actually represent three different sources of merged MBHs: 1) the leftmost branch is the ``light population," composed by two \acp{MBH}, both still having roughly the initial seed mass. 
This commonly happens at high redshift and in small dark matter halos at low redshift, due to the lack of previous merger. 
Without mergers, quiet accretion of seed black holes alone is quite inefficient. 
Thus, \acp{MBH} keep almost the initial seed mass till the first merger happens. 
The light population is approximately $10^5 \Msun$ for the H-seed model and approximately $10^2 \Msun$ for the L-seed model. 
2) In the medium population, one of the \acp{MBH} has not yet accreted from the initial seed mass, while the other one has already experienced a merger and thus has evolved into a \ac{MBH} with approximately $10^{7}-10^{8} \Msun$, making the medium chirp mass population approximately $10^6 \Msun$ for the H-seed model and approximately $10^4 \Msun$ for the L-seed model. 
3) In the heavy population, both \acp{MBH} have already evolved. 
The heavy populations of both the H-seed and L-seed models is composed of systems with chirp mass approximately $10^7 \Msun$.

The work presented in Ref. \cite{EAGLE2016} presents a meaningful comparison to the H-seed model. 
The EAGLE simulation adopted hydrodynamical simulation for the galaxy evolution, and the dark matter halo evolution was also obtained through N-body simulation. 
In Fig. 4 of Ref. \cite{EAGLE2016}, the three population of mergers were also identified.
Unfortunately, there is no similar comparison for the L-seed model. 

Note that the division in mass in Fig. \ref{fig_cm_red_SNR} could originate from a combination of specific physical assumptions and limited Millennium-I resolution.
In the adopted \ac{MBH} accretion model, fast mass growth (quasar mode) is only triggered by mergers. 
Moreover, seed black holes are placed in each halo as soon as it appears in the simulation. 
This means that, by construction, the first generation of mergers involves \acp{MBH} with masses close to the initial seed mass. 
Galaxies that  have not experienced any major merger are still very gas rich, and cold gas could be as large as a few percent of the host halo virial mass. 
During each merger event, approximately $0.1\% - 1\%$ of cold gas is accreted into the central \ac{MBH}. 
We emphasize that the halo resolution of Millennium-I simulation is $2.4\times 10^{10}\Msun$. 
A wet (or gas rich) merger in such a massive halo can feed seed black holes to approximately $10^7 \Msun$ through just one or few merger events. 
This rapid growth of seed black holes allows them to ``catch up" with the massive halo mass, but also causes the gap between seed black holes ($10^2 \Msun$ or $10^5 \Msun$) and evolved \acp{MBH} (approximately $10^7-10^8\Msun$, no matter what the seed mass is, which is also due to the rapid growth). 
The separation into three distinct populations is therefore potentially artificial. 
Allowing \ac{MBH} growth via secular processes not driven by mergers would widen the spectrum of masses involved in the first merger events, blending the three populations. 
Moreover, smaller seed black holes could appear early at $z\sim 15-20$, while Millennium-I simulation set an upper limit of redshift of $z\sim 12$.
By this time, low-mass seeds formed at higher redshift could have already undergone several merger events, resulting in a much higher mass. 
An increase of the resolution of the dark matter simulation and different assumptions in the triggering of \ac{MBH} accretion would potentially result in a blending of the three subpopulations in a continuous distribution.
However, that is beyond the scope of this work, and we leave it for future discussion.
 
\subsection{EPS-based models}

Besides the Millennium-I-based models described in the previous section, we also consider a different set of \ac{MBH} populations, produced with the semianalytic galaxy formation model of Ref. \cite{EB12} (with incremental improvements described in Refs. \citep{Sesana14,Antonini_long}).
The EPS formalism was calibrated in which it can reproduce properties of the Millenium-I simulation.
It has the advantage of a higher mass-resolution and thus theoretically provides a complete description of \ac{MBH} binary mergers, the cost of which is that it adopts some {\it ad hoc} assumptions for the dark matter halo evolution.

The EPS model adopted in this work was also extensively used to assess the expected LISA scientific performance as a function of experimental design, focusing in particular on the physics of MBH mergers \cite{Klein16,Bonetti_triple} and their electromagnetic counterparts \cite{Tamanini2016}, and also on extreme mass-ratio inspirals \citep{Babak2017} and MBH-based ringdown tests of general relativity (GR) \cite{ringdown}. 
Moreover, the model was also used to predict the stochastic \ac{GW} background expected for pulsar timing array experiments \cite{Dvorkin2017,Bonetti2018b}.

We adopt here the same model as in Refs. \cite{Klein16,Antonini_long}.
The evolution of dark matter halos is followed via an EPS formalism \cite{1974ApJ...187..425P}, suitably tuned to reproduce the results of N-body simulations \cite{Parkinson2008}. 
On top of this dark matter skeleton, baryonic structures are evolved semianalytically. 
In particular, such structures include a chemically pristine intergalactic medium, which either streams into the dark matter halos via cold flows \cite{cold_flows} or is shock heated to the halo's virial temperature and then cools down adiabatically.
The end result of both processes is the formation of a cold gas component that can give rise to star formation (with ensuing supernova explosions and feedback on the formation of stars itself), which in turn chemically enriches the gas.
Both these cold gas and stellar components may exist in disks or bulges, with the latter being produced by the disruption of the disks as a result of major galactic mergers or bar instabilities. 
Both major mergers and bar instabilities are expected to drive an excess of cold chemically enriched gas to the nuclear region of the galaxy. 
We model this gas transfer by assuming its rate is proportional to star formation in the bulge \cite{Granato2004,Haiman2004,Lapi2014}.
The nuclear gas ``reservoir" that forms as a result is then assumed to either accrete onto the central \ac{MBH} on a viscous timescale or form stars {\it in situ}, giving rise to nuclear star clusters \cite{Antonini_long,Antonini_short}. 
The latter also grow as a consequence of dynamical friction driven migration of globular clusters from the galaxy's outskirts to the center \cite{Antonini_long,Antonini_short}.
AGN feedback is included in the model, and with the aforementioned prescriptions for \ac{MBH} growth by accretion from the nuclear reservoir, it ensures that local observed correlations between \ac{MBH} and galaxy properties are reproduced at $z\sim 0$ \citep{EB12,Sesana14,Barausse2017}.

As mentioned, the mergers of the dark matter halos are followed via the underlying merger tree. 
The EPS-based model first accounts for the delay between the time the halos first touch (as extracted from the merger tree) and the time the two halos and the hosted galaxies finally merge, by using the results of Ref. \cite{Boylan-Kolchin2008}. 
We also account for the tidal disruption and heating of the satellite halo (and galaxy) by following Ref. \cite{Taffoni2003}. 
We then account for the delay between the galaxy merger and the final \ac{MBH} merger.
There is considerable uncertainty about this timescale, with suggestions that it may even exceed the Hubble time in some cases (this is the so-called final parsec problem; see, e.g. Ref. \cite{Colpi2014} for a review). 
In the following, we follow Refs. \cite{Antonini_long,Klein16} and adopt different timescales according to the environment of the \ac{MBH} binary, with delays of a few gigayears when the binary is driven to coalescence by stellar hardening and of the order of approximately $10^7$--$10^8$ yr when the \ac{MBH} binary shrinks as a result of migration in a nuclear gas disk.
We also account for the effect of triple \ac{MBH} interactions on these delay times in a simplified way, described in Ref. \cite{Antonini_long}, to which we refer, more in general, for a more detailed description of the model for these delay times. 
(See also Refs. \cite{Bonetti2016,Bonetti2017,Bonetti2018a,Bonetti2018b,Ryu2017,Bonetti_triple} for more recent work on triple \ac{MBH} interactions).
To assess the impact of the aforementioned uncertainties on the physics of the delays between galaxy and \ac{MBH} mergers, we also consider models where these delays are set to zero. 
(However, we keep the delays between halo contact and galaxy/halo mergers, modeled as described above.)

As for the high redshift seeds of the \ac{MBH} population, we adopt, as in Refs. \cite{Antonini_long,Klein16}, a light-seed model whereby \ac{MBH} seeds form as remnants of popIII stars\ (thus the name {popIII}) \cite{2001ApJ...551L..27M} and a heavy-seed one where they originate from bar-instabilities of protogalactic disks \cite{Volonteri2008}.
We adopt $Q_c=3$ for the critical Toomre parameter for disk instability, thus name the heavy seed models as {Q3}.
In the former model, the typical \ac{MBH} seed is approximately $100 M_\odot$, while in the latter the seed mass is of the order of approximately $10^5 M_\odot$.
Following Ref. \cite{2014ApJ...784L..38M}, we allow for moderately super-Eddington accretion in the light-seed models to ease the discrepancy between that model and observations quasars at $z\sim6$--7. 
We refer again to Refs. \citep{Antonini_long,Klein16} for more details on the seed model.

\begin{figure}[htb]
\centering
\begin{subfigure}[b]{.5\linewidth}
\centering
\includegraphics[width=\textwidth,height=7cm]{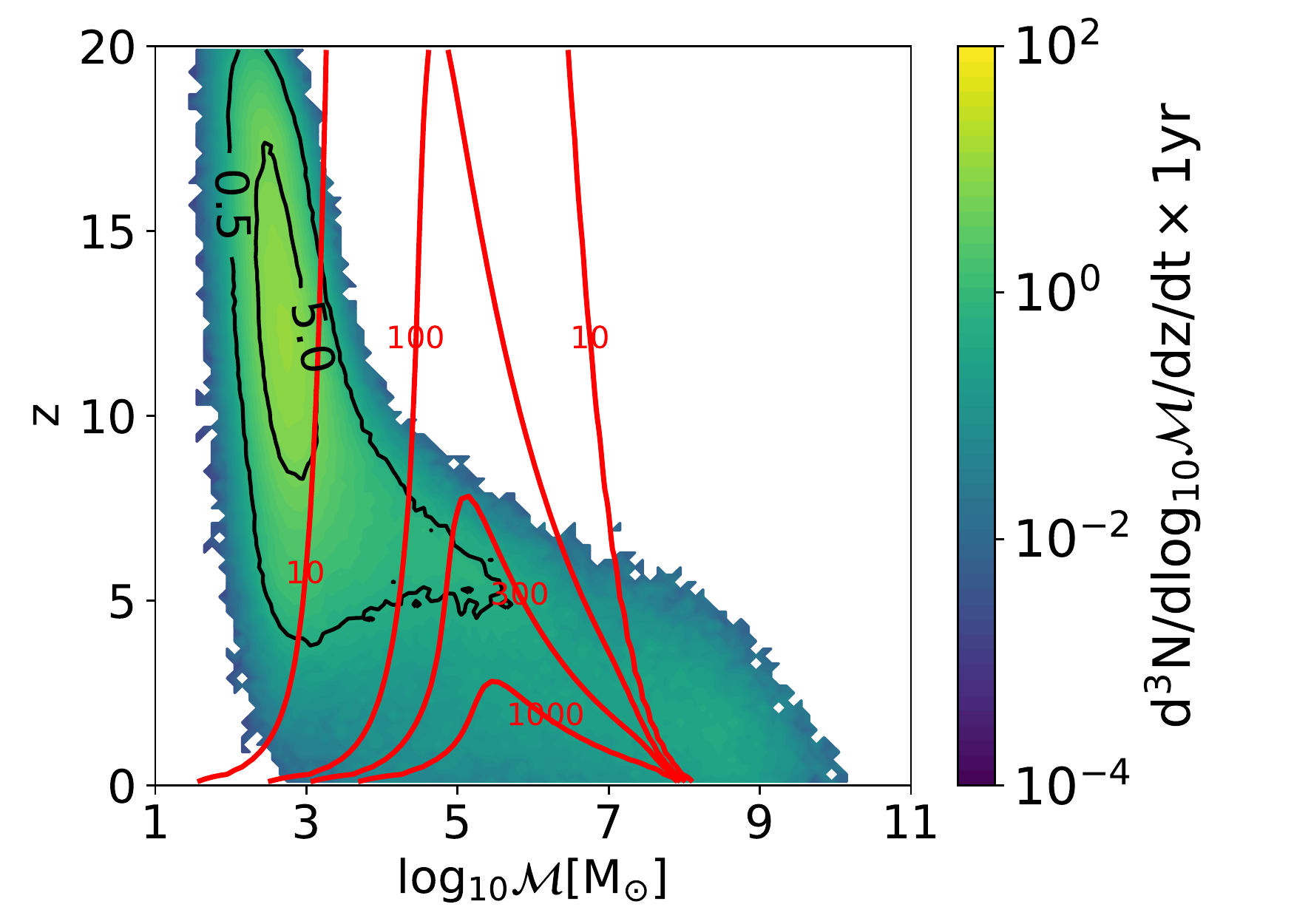}
\caption{popIII}\label{A.Klein distributions1}
\end{subfigure} \\%
\begin{subfigure}[b]{.5\linewidth}
\centering
\includegraphics[width=\textwidth,height=7cm]{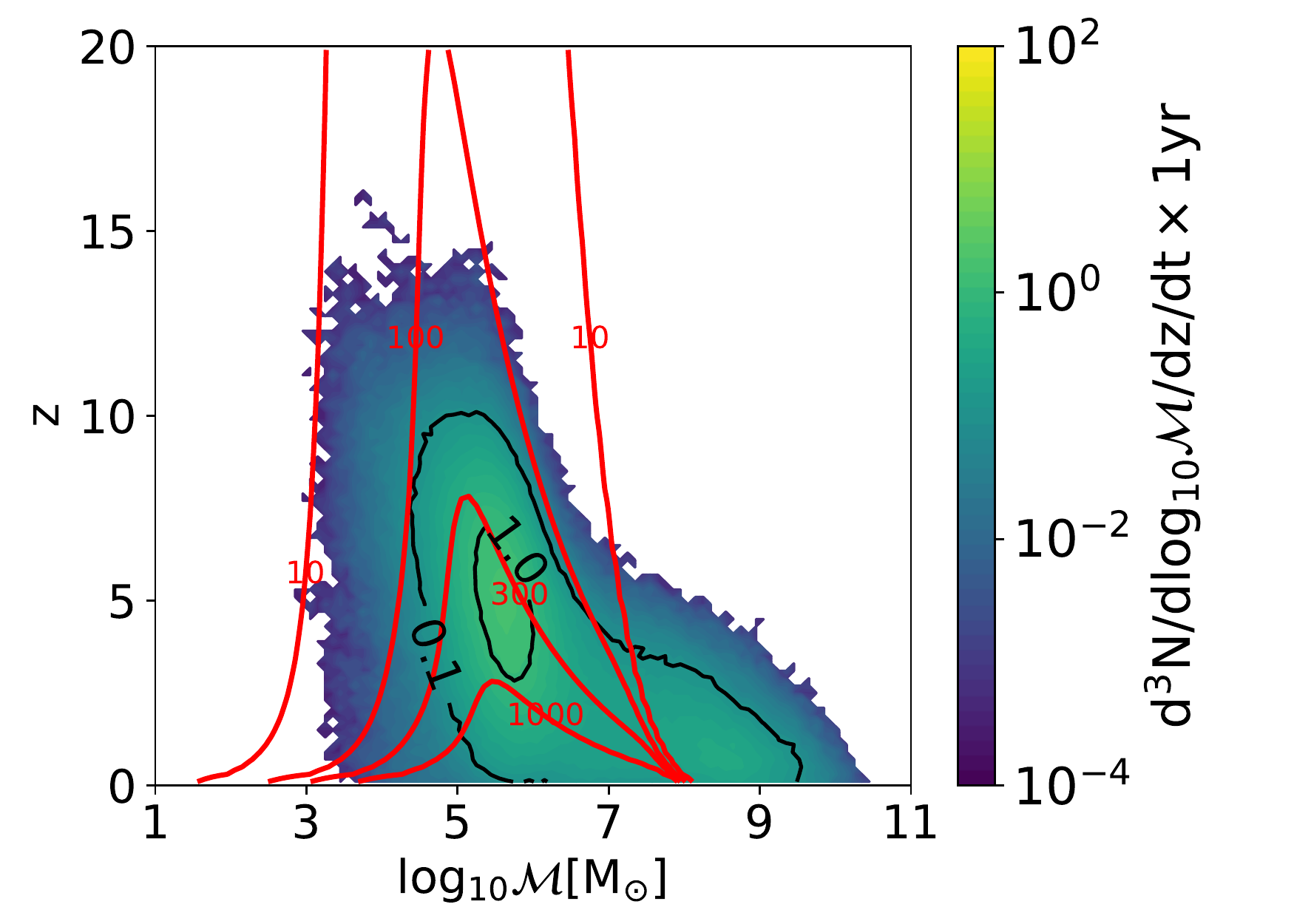}
\caption{Q3\_d}\label{A.Klein distributions2}
\end{subfigure}%
\begin{subfigure}[b]{.5\linewidth}
\centering
\includegraphics[width=\textwidth,height=7cm]{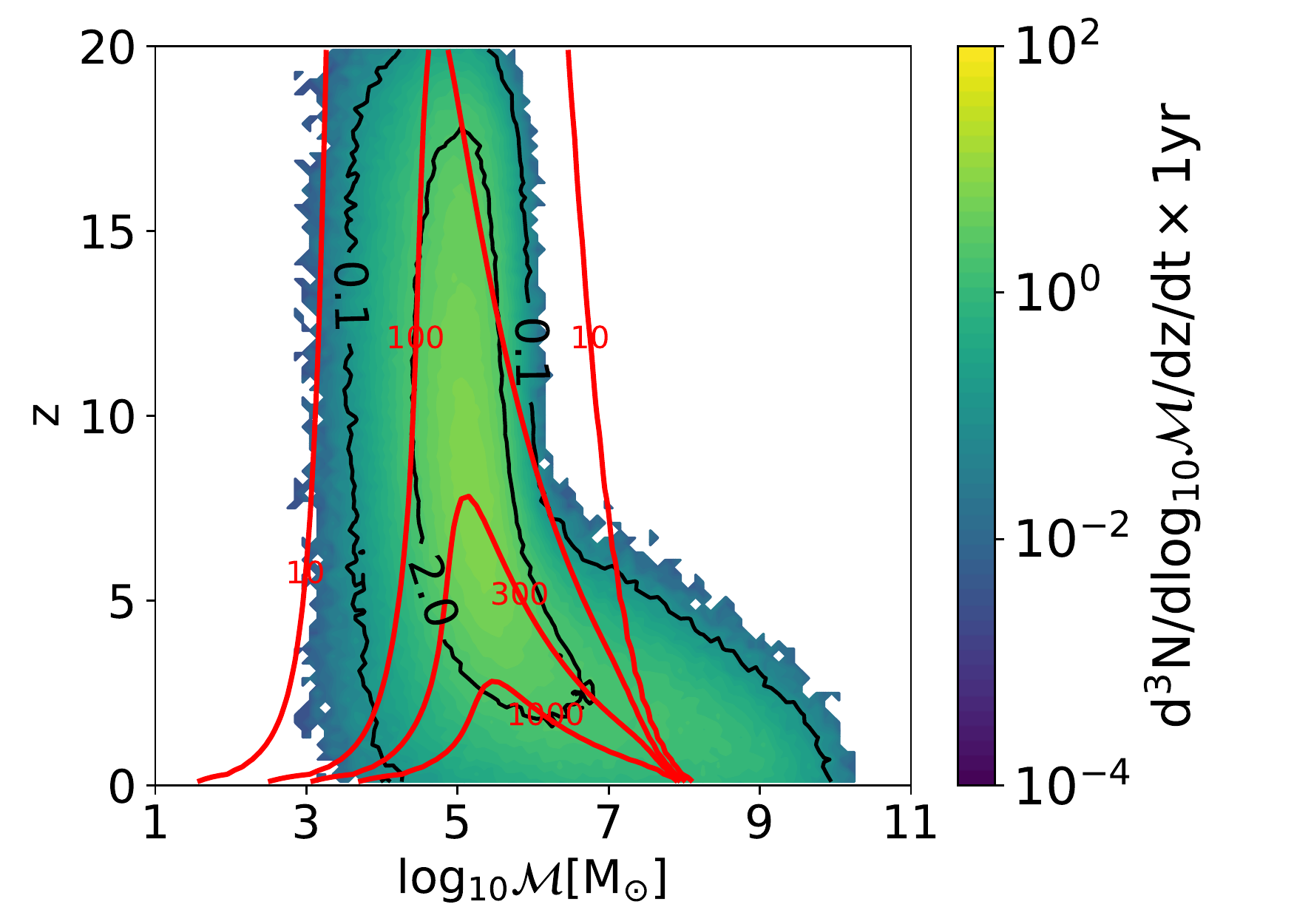}
\caption{Q3\_nod}\label{A.Klein distributions3}
\end{subfigure}%
\caption{The distribution of mergers over chirp mass and redshift of the popIII model(upper panel), Q3\_d model(lower left panel) and Q3\_nod model(lower right panel).
The red contour line represents the average SNR in the TianQin detector, assuming equal-mass binaries (details of calculation are discussed in Sec. \ref{sec:SNR}). 
 The black lines represent contours on the differential number of mergers throughout the cosmic history.
}
\label{A.Klein distributions}
\end{figure}

The predicted merger distributions of the models are shown in Fig. \ref{A.Klein distributions}. 
The higher resolution of the EPS merger trees, together with the different prescription for \ac{MBH} evolution, results in unimodal distributions in all cases. 
The time delay does not impact hugely for the popIII model, as pointed out in Ref. \cite{Klein16}; thus, we only study the time-delayed version of the model.
The influence of including delays in the \ac{MBH} binary mergers is apparent by comparing the Q3\_d and Q3\_nod panels. 
While in the latter mergers already occur at $z\sim20$, in the former they start to occur only at $z<15$.

Notice the Millennium-I simulation has a redshift upper limit of $12$, while the EPS models can depict mergers with redshift as high as $20$.
Also, the evolutionary path of the mergers is more evident in Fig. \ref{A.Klein distributions}; the \acp{MBH} most likely to merge demonstrate the tendency of increasing mass as the Universe evolves.
It is also interesting to notice that the three different models demonstrate very different distribution in terms of mass over redshift.
The popIII model predicts the most detectable source origin at a redshift of $5-10$, with masses lean in the low-mass end; the Q3\_d model predicts the most mergers happen at redshift $z<10$, and the Q3\_nod model predicts a wide spread of redshift for mergers.

%%%%%%%%%%%%%%%%%%%%%%%%%%%%%%%%%%%%%%%%%%%%%%%%%%%%%%%%%%%%%%%%%%%
\section{Key quantities and method}\label{sec:method}
Having described the \ac{MBH} population models, we now turn our attention on the \ac{GW} part. 
In the following, we describe the mathematical tools adopted to define source detection and parameter estimation accuracy, the sensitivity of the TianQin detector, as well as its response function to an incoming \ac{GW} as a function of time.

%%%%%%%%%%%%%%%%%%%%%%%%%%%%%%%%%%%%%%%%%%%%%%%%%%%%%%%%%%%%%%%%%%%
\subsection{SNR and parameter estimation}

% introduce inner product
In \ac{GW} data analysis, it is useful to define the inner product between two waveforms ${\tilde h_1}(f)$ and ${\tilde h_2}(f)$ as $(h_1|h_2)$, 
\begin{equation}
%\begin{aligned}
  (h_1{\vert}h_2) \equiv 2\int_{f_{low}}^{f_{max}} \frac{{\tilde h_1}^*(f){\tilde h_2}(f) + {\tilde h_2}^*(f){\tilde h_1}(f)}{S_n(f)} {\rm d}f,
%\end{aligned}
\label{eq:innerp}
\end{equation}
where the choice of $f_{low}\geq 0$ and $f_{max}$ is case dependent and $S_n(f)$ is the \ac{PSD} for the detector.

The \ac{SNR} $\rho$ of a signal can be then expressed as
\begin{eqnarray}
\rho \equiv (h{\vert}h)^{1/2}.
\label{eq:rho}
\end{eqnarray}
For multiple detectors, the combined \ac{SNR} is simply the root sum square of component \acp{SNR}.

% definition FIM
The \ac{FIM} is often used to quantify the uncertainty in the estimation of the relevant waveform parameters as well as their mutual correlations. 
The \ac{FIM} $\Gamma_{ij}$ is defined as
\begin{align}
\Gamma_{ij}\equiv\left(\frac{\partial h}{\partial\theta^i}\middle|\frac{\partial h}{\partial\theta^j}\right)
\end{align}
The variance-covariance matrix $\Sigma_{ij}$ is related to the \ac{FIM} $\Gamma_{ij}$ through $\Sigma_{ij} = \Gamma_{ij}^{-1}$.
The uncertainty of a given parameter $\theta_i$ is then $\Delta\theta^i=\sqrt{\Sigma^{ii}}$, and the correlation coefficients between any two parameters $\theta^i$ and $\theta^j$ is
$$c_{ij}=\Sigma^{ij}/\sqrt{\Sigma^{ii}\Sigma^{jj}}.$$
In calculating the sky localization error $\Delta\Omega$, we adopt the formula
\begin{eqnarray}
\Delta\Omega=2\pi |sin\theta|(\Sigma_{\theta\theta}\Sigma_{\phi\phi}-\Sigma_{\theta\phi})^{1/2}
\end{eqnarray}\label{eq:omega0}

For post-Newtonian (PN) inspiral waveforms, most of the partial derivatives $\frac{\partial h}{\partial\theta^i}$ can be analytically obtained. 
However, the merger and ringdown phase contains a certain portion of the total \ac{SNR} of merging \ac{MBH} binaries. 
Therefore, a waveform comprehensively describing the whole inspiral-merger-ringdown process is better suited to our investigation. 
In the following, we therefore adopt the IMRPhenomP waveforms \cite{PhysRevLett.113.151101} throughout our analysis. 
Notice that by adopting IMRPhenomP waveforms, we cannot easily obtain closed-form partial derivatives.
We therefore approximate partial derivative numerically through numerical differentiation: 
\begin{eqnarray}\label{eq:delta0}
  \frac{\partial h}{\partial\theta_i} \approx  \frac{\Delta h}{\Delta\theta_i}\equiv \frac{h(\theta_i+\Delta\theta_i)-h(\theta_i)}{(\theta_i+\Delta\theta_i)-\theta_i}\,.
\end{eqnarray}
The specific value of $\Delta \theta_i$ is chosen once the corresponding $\Gamma^{ii}$ reaches convergence.
If available, we also compare numerical differentiation results with analytical results with PN inspiral waveforms for a correctness check.

In the actual analysis, we perform calculation over the nine parameters, chirp mass $\mathcal{M}$, symmetric mass-ratio $\eta=(m_1m_2)/(m_1+m_2)^2$, luminosity distance $D_L$, merger phase $\phi_c$, merger time $t_c$, location angles $\theta$ and $\phi$, and spin of two black holes $\chi_1$ and $\chi_2$.
Notice that we do not include the inclination angle $\iota$ nor the polarisation angle $\psi$ in our analysis, since adding those parameters would introduce degeneracies that make the inversion of \ac{FIM} problematic. 

In a different work that investigates the parameter estimation accuracy from MBH binary inspiral signals, third order restricted PN waveform including nonprecession spin and first order eccentricity effects has been used for a single Michelson interferometer of TianQin \citep{Feng2019}. 

\subsection{Gravitational wave signal in the detector}

To describe the response of the detector to the incoming \ac{GW}, it is convenient to discuss it in the detector coordinate system.
For a source located at longitudinal angle $\theta_S$ and azimuthal angle $\phi_S$, with polarization angle $\psi_S\,$, the signal detected in TianQin is
\begin{eqnarray}\label{eq:h0}
%\begin{aligned}
  h_{\mathrm {I,II}}(t)=h_+(t)F_{\mathrm {I,II}}^+\big(\theta_S,\phi_S,\psi_S\big)+h_{\times}(t)F_{\mathrm {I,II}}^{\times}\big(\theta_S,\phi_S,\psi_S\big)\,,
%\end{aligned}
\end{eqnarray}
% 1 TianQin = 2 orthogonal Michelson
where the subscripts I and II correspond to the two equivalent orthogonal Michelson interferometers.
And we have 
\begin{eqnarray}
%\begin{aligned}
h_+(t)=h_0(t)(1+\nu^2)/2\,,\quad h_{\times}(t)=h_0(t)(-i\nu)\,,
%\end{aligned}
\end{eqnarray}
$\nu=\cos \iota=\hat{L}\cdot \hat{n}\,$, with $\iota\,$ being the inclination angle.
$F_{+,{\times}}$ are antenna pattern functions, which describe the detector response to sources with different locations and polarizations.
The antenna pattern is frequency dependent, but in the low frequency limit, which holds true for most of our analysis, it can be simplified as
\begin{eqnarray}\label{eq:f1}
F_\mathrm{I}^+\big(\theta_S,\phi_S,\psi_S\big)&=&\frac{\sqrt3}2\Big[\frac{1}{2}(1+\cos^2\theta_S)\cos2\phi_S\cos2\psi_S -\cos \theta_S \sin2 \phi_S \sin2 \psi_S\Big]\,,\nn\\
F_\mathrm{I}^{\times}\big(\theta_S,\phi_S,\psi_S\big)&=&\frac{\sqrt3}2\Big[\frac{1}{2}(1+\cos^2\theta_S)\cos2\phi_S\sin2 \psi_S + \cos \theta_S \sin2\phi_S\cos2\psi_S\Big]\,,
\end{eqnarray}
and the antenna pattern function of the second orthogonal Michelson interferometer can be written as
\begin{eqnarray}\label{eq:f2}
\begin{aligned}
F_{\mathrm{II}}^+\big(\theta_S,\phi_S,\psi_S\big) = F_\mathrm{I}^+\big(\theta_S,\phi_S-\frac{\pi}{4},\psi_S\big)\,,\\
F_{\mathrm{II}}^{\times}\big(\theta_S,\phi_S,\psi_S\big) = F_\mathrm{I}^{\times}\big(\theta_S,\phi_S-\frac{\pi}{4},\psi_S\big)\,.
\end{aligned}
\end{eqnarray}
For higher frequencies where the low frequency approximation lost validity, the impact and corresponding treatment is discussed in Sec. \ref{sec:sensitivity}. 

The polarization angle $\psi_S$ can be expressed as
\begin{eqnarray}\label{eq:psi}
\tan\psi_S = \frac{\hat{L}\cdot\hat{z}-(\hat{L}\cdot\hat{z})(\hat{z}\cdot\hat{n})}{\hat{n}\cdot(\hat{L} \times{\hat{z}})}
\end{eqnarray}
where $\hat{L}$ and $-\hat{n}$ are the unit vector along the orbital angular momentum and the direction of GW propagation, respectively. 
Since the plane of the TianQin constellation is nearly fixed in space \cite{Luo:2015ght}, both $\theta_S$ and $\psi_S$ are nearly time independent, while $\phi_S$ is linearly proportional to operational time. 
Thus, one can derive the detected signal as the convolution between the antenna pattern and the frequency domain waveform,
\begin{eqnarray}\label{eq:h1h2}
  {\tilde h_\mathrm{I}}(f)&=&\mathscr{F}\{h_\mathrm{I}(t)\}={\tilde h_\mathrm{I}^+}(f)+{\tilde h_\mathrm{I}^{\times}}(f) \nn\\
  {\tilde h_\mathrm{I}^+}(f)&=&F_\mathrm{I}^+\big(\theta_s,\phi_s(f),\psi_s\big){\circledast}{\tilde h^+}(f), \nn\\
  {\tilde h_\mathrm{I}^{\times}(f)}&=&F_\mathrm{I}^{\times}\big(\theta_s,\phi_s(f),\psi_s\big){\circledast}{\tilde h^{\times}}(f),
\end{eqnarray}
where $\mathscr{F}\{h_\mathrm{I}(t)\}$ means Fourier transformation of $h_\mathrm{I}(t)\,$, and
\begin{eqnarray}\label{eq:h1h2f}
  {\tilde h_\mathrm{I}^+}(f)&=&\frac{1}{4}(1+\cos^2\theta_s)\cos2\psi_s \left\{{\tilde h^+}(f+2f_0)e^{-i2\phi}+{\tilde h^+}(f-2f_0)e^{i2\phi}\right\} \nn\\
  &-&\frac{i}{2}\cos\theta_s\sin2\psi_s\left\{{\tilde h^+}(f+2f_0)e^{-i2\phi}\right. -\left.{\tilde h^+}(f-2f_0)e^{i2\phi}\right\}, \nn\\
  {\tilde h_\mathrm{I}^{\times}}(f)&=&\frac{1}{4}(1+\cos^2\theta_s)\sin2\psi_s \left\{{\tilde h^{\times}}(f+2f_0)e^{-i2\phi}+{\tilde h^{\times}}(f-2f_0)e^{i2\phi}\right\} \nn\\
  &+&\frac{i}{2}\cos\theta_s\cos2\psi_s\left\{{\tilde h^{\times}}(f+2f_0)e^{-i2\phi}\right. -\left.{\tilde h^{\times}}(f-2f_0)e^{i2\phi}\right\}\,.
\end{eqnarray}

$h_{\mathrm {II}}$ has analogue expressions with an extra $\phi$ rotation of $\pi/4$.

The annual orbit of the Earth further introduces a Doppler correction $h(f)e^{-i\varphi_D(t(f))}$,
where $t(f)$ is approximated  under 0PN $t_c-\frac{5}{256\mathcal{M}_z}(\pi\mathcal{M}_zf)^{-8/3}$, $\mathcal{M}_z=\mathcal{M}(1+z)$ is the redshifted chirp mass.
And $\varphi_D(t)$ is given by
\begin{eqnarray}\label{eq:doppler}
\varphi_D(f)=\frac{2\pi f}{c}R \sin\bar{\theta}_S \cos\Big(\bar{\phi}\big(t(f)\big)-\bar{\phi}_S\Big), 
\end{eqnarray}
where $R=1 AU$ and $\bar{\phi}(t)=\bar{\phi}_0+2\pi t/T$,
and $\bar{\phi}_0$ specifies the detector's location at $t=0$. The angles $(\bar{\theta}_S,\bar{\phi}_S,\bar{\phi}_0)$ are the locations of detectors relative to the Sun.
$T=1$ year is the orbital period of TianQin.

The gravitational evolution of a binary black hole system can be roughly divided into three stages: inspiral, merger and ringdown. 
In the inspiral stage, the two black holes are well separated, so \ac{PN} expansion is sufficient to describe the system to high accuracy \cite{2014LRR....17....2B}. 
The merger stage is relatively short but complex, so \ac{NR} is required to depict the details \cite{2010nure.book.....B} (see, however, Ref. \citep{2018arXiv181000040M}). 
The ringdown stage can be understood through perturbation theory of Kerr black holes.

For the generation of waveform, we adopt a self-consistent waveform family that contains the whole inspiral-merger-ringdown stages, known as IMRPhenomPv2 \cite{PhysRevLett.113.151101}. 
This waveform model is implemented in the LIGO Algorithm Library \cite{lal}. 
It uses an approximate waveform of precessing black hole binaries calibrated with \ac{PN} and \ac{NR}. 
The waveform model was adopted for LIGO detections of stellar-mass black holes, but under proper modification of $h(f|\alpha\mathcal{M})=\alpha^2 h(f/\alpha\,|\mathcal{M})$, it could also be used for MBH systems. 

Using the \ac{PN} approximation, one can derive a lower boundary on the frequencies \cite{1994PhRvD..49.2658C}:
\begin{equation}
\begin{aligned}
f_{low}&=(256/5)^{3/8}\frac{1}{\pi}{\mathcal{M}_z}^{-5/8}(t_c-t)^{-3/8}\,.
\end{aligned}
\label{eq:f_low}
\end{equation}
This result, together with $f_{max}=\infty$, will be used for the evaluation of the inner product in Eq. (\ref{eq:innerp}).

\subsection{TianQin sensitivity}\label{sec:sensitivity}

\begin{figure}
\includegraphics[width=0.7\textwidth]{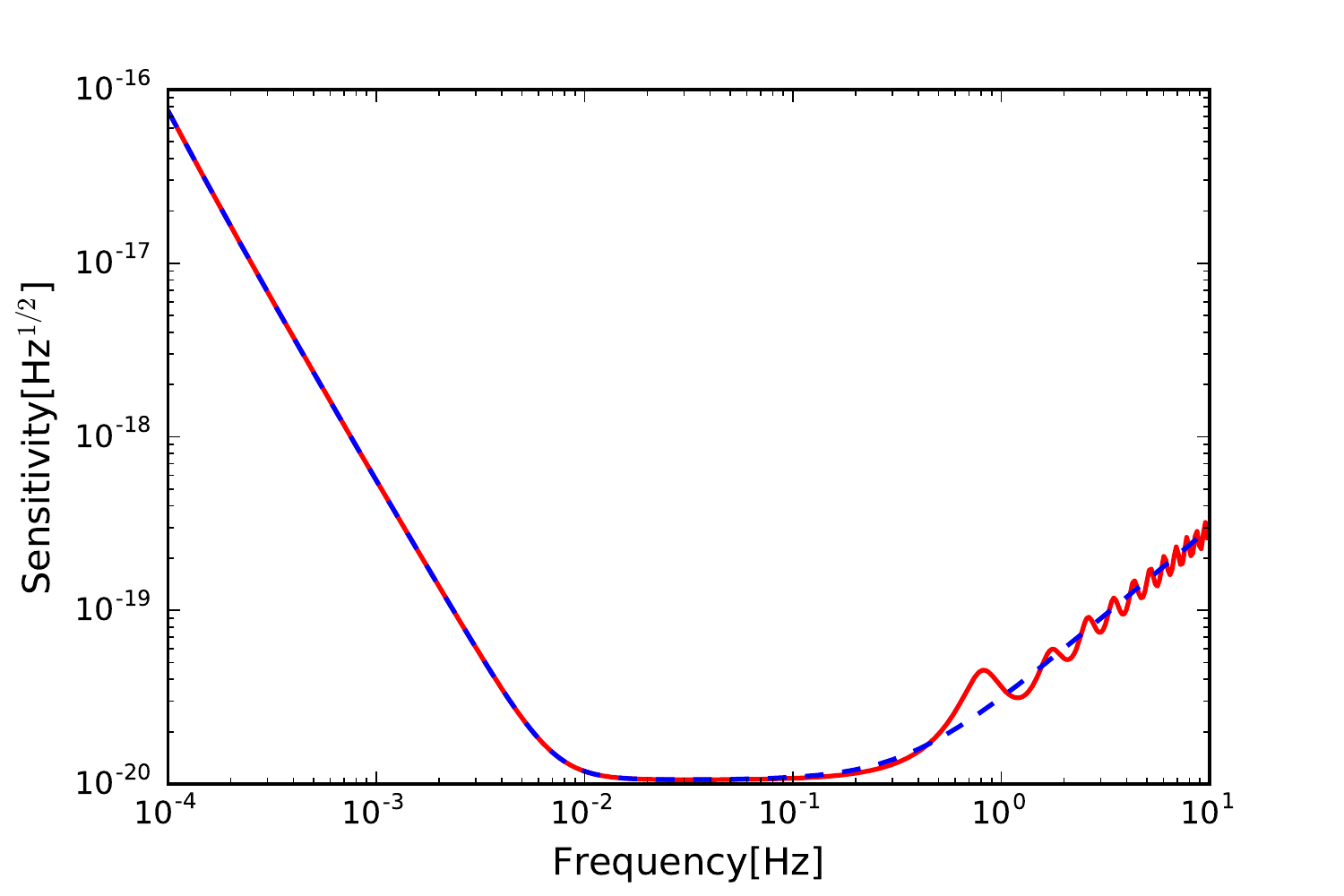}
\caption{Anticipated sensitivity curve for TianQin. The red solid line corresponds to Eq. (\ref{eq:heffbar}), while the dashed blue line corresponds to taking $g(x)\approx1\,$.}
\label{fig:ASD}
\end{figure}

In this paper, we adopt the following model for the sky averaged sensitivity of TianQin \citep{Luo:2015ght}, 
\begin{eqnarray}
%\begin{aligned}
  &&S_n^{SA}(f)=\frac{S_N(f)}{\overline{R}(2\pi f)}\,,\nn\\
&&S_N(f)=\frac1{L^2}\left[\frac{4S_a}{(2\pi f)^4}\Big(1+\frac{10^{-4}{\rm Hz}}f\Big)+S_x\right]\,,\nn\\
&&\overline{R}(w)=\frac{3}{10}\times\frac{g(w\tau)}{1+0.6(w\tau)^2}\,,
%\end{aligned}
\label{eq:heffbar}
\end{eqnarray}
where $S_a^{1/2}= 1\times 10^{-15}\Sa$, $S_x^{1/2} = 1\times 10^{-12}\Sx$, $\tau=L/c$ is the light travel time for a TianQin arm length, and
\begin{eqnarray}
%\begin{aligned}
g(x)=\left\{\begin{matrix}\sum_{i=0}^{11}a_ix^i&:&x<4.1\,,\cr\cr
\exp\left[-0.322\,\sin(2x-4.712)+0.078\right]&:&4.1\leq x<\frac{20\pi}{\sqrt3}\,,\end{matrix}\right.
%\end{aligned}
\label{eq:gx}
\end{eqnarray}
with the coefficients $a_i$ given in Table \ref{tab:coefficients-ai}. 
The unusual expression for $g(x)$ is an analytical fit to numerical calculations, and the upper limit on $x$ in the function $g(x)$ corresponds to $f=10$Hz. 
Both the full result and the further approximation with $g(x)\approx1$ is illustrated in Fig. \ref{fig:ASD}. 
Note that the expression of Eq. (\ref{eq:heffbar}) agrees with our previous  study  \citep{2018CQGra..35i5008H} in which Monte Carlo simulation has been adopted to evaluate the sky averaged sensitivity curve.

\begin{table}
\begin{center}
\begin{tabular}{|c|c|c|c|c|c|}
  \hline
  $a_0$ & $a_1$ & $a_2$ & $a_3$ & $a_4$ & $a_5$\\
  \hline
  &&&&&\\
  1 & $\displaystyle\frac1{10^4}$ & $\displaystyle\frac{2639}{10^4}$ & $\displaystyle\frac{231}{5\times10^4}$ & $-\displaystyle\frac{2093}{1.25\times10^4}$ & $\displaystyle\frac{2173}{10^5}$ \\
  &&&&&\\
  \hline
  $a_6$ & $a_7$ & $a_8$ & $a_9$ & $a_{10}$ & $a_{11}$\\
  \hline
  &&&&&\\
  $\displaystyle\frac{2101}{10^6}$ & $\displaystyle\frac{3027}{2\times10^5}$ & $-\displaystyle\frac{42373}{5\times10^6}$ & $\displaystyle\frac{176087}{10^8}$ & $-\displaystyle\frac{8023}{5\times10^7}$ & $\displaystyle\frac{5169}{10^9}$ \\
  &&&&&\\
  \hline
\end{tabular}
\caption{Coefficients of Eq. (\ref{eq:gx}) for the response of TianQin to a signal.}
\label{tab:coefficients-ai}
\end{center}
\end{table}

The low frequency behavior of the acceleration noise for TianQin is not clear yet. 
We thus adopt a conservative lower frequency cutoff at $10^{-4}$Hz, and this effectively sets the noise PSD under this frequency to be infinity. 
As a consequence, the conclusions in this paper should also be taken to be conservative in this respect.

Due to its particular choice of orbit, TianQin adopts a ``3 month on + 3 month off" observation scheme. 
It is therefore interesting to consider the scenario when twin set of TianQin constellations operate consecutively, filling up the observation gaps for each other. 
We note such a scheme will not modify the sensitivity curve for TianQin.

%%%%%%%%%%%%%%%%%%%%%%%%%%%%%%%%%%%%%%%%%%%%%%%%%%%%%%%%%%%%%%%%%
\section{Results}\label{sec:result}

\subsection{SNR}\label{sec:SNR}

% Describe fig 1/2 and the physical meaning
In Figs. \ref{fig_cm_red_SNR} and \ref{A.Klein distributions}, we plot the distribution of mergers over chirp mass and redshift from the models in Ref. \cite{Jiang2018} as well as in Ref. \cite{Klein16}. 
Throughout the whole simulated Universe, all mergers happening at different stages were recorded and binned.
Each tile of chirp mass and redshift was assigned with a merger density of $\frac{{\rm d}^3N }{{\rm d}\log_{10}\mathcal{M}\, {\rm d}z\,{\rm d}V_c} $ with respect to comoving volume. 

% SNR plot
Overplotted in red are contour plots of \acp{SNR} in the TianQin detector, assuming equal-mass binaries and a fiducial observation time of three months before merger.  
For most of the \ac{MBH} merger events, the majority of the \ac{SNR} comes from the last weeks or even days before final merger, so this arbitrary choice of observation time should not incur too much error on the estimated \ac{SNR} \cite{Feng2019}.

% 3month + 3month
For the sake of accuracy, when applied for the calculation of the detection rate in Sec. \ref{sec:rate}, we do include the true operation scenario of TianQin.
The ``3 month on + 3 month off" working pattern introduces a minor complication for the \ac{SNR} calculation.
The calculation adopted is actually 
\begin{equation}\label{eq:snr3m}
  \rho^2 = 4\sum_i \int_{f_{lo}^i}^{f_{hi}^i} {\rm d}f \frac{h(f)h^*(f)}{S_n(f)}
\end{equation}
by randomly assigning merger time $t_c \in U[0,5]$yr and fixing the starting and ending times for each three-month session (for example, $0$ and $ 3$  months, respectively), one can determine $f_{lo}^i$ and $f_{hi}^i$ through Eq. (\ref{eq:f_low}).

\subsection{Detection rate}\label{sec:rate}
%%%%%%%%%%%%%%%%%%%%%%%%%%%%%%%%%%%%%%%%%%%%%%%%%%%%%%%%%%%%%%%%%
We adopt a conventional choice \ac{SNR} threshold of $8$ for detection. 
Note a conservative lower frequency cutoff in $10^{-4}Hz$ is used when computing \acp{SNR}; thus, the detection results, especially the ability for high-mass events, should be also regarded as conservative.

In all models, we simulate a random realization of merger catalogs for within a nominal five year operation time. 
The physical parameters including redshift, mass, etc., were obtained from the models, while the merger times were assigned uniformed within observation time.
%The \ac{SNR} accumulated long before merger is negligible compared with the last several days, for the sake of convenience, we do not include events merge after the operation ends.
The SNR of a MBH binary coalescence is dominated by the final days before merger, so binaries merging after the operation end are not considered in the calculation of the detection rate, for the sake of convenience as well as conservativeness. 
We then simply count how many events out of the catalogue would induce an \ac{SNR} larger than a threshold of $8$, and by averaging over multiple trials, one can obtain the expected detection rate.
Notice that we also perform detection rate calculation similar to Ref. \cite{EAGLE2016} for correctness check, and the results are quite consistent.

% comparison of one TianQin and twin TianQin
It is meaningful to consider a {twin constellations} scenario for TianQin, where two sets of constellation relay the \ac{GW} detections. 
In this configuration, the perpendicular twin constellations would have a complete time coverage of all events, essentially double the detection rate.

% showing comparison
In Table \ref{event_rate_all},
%we summarize the detection rates with TianQin for all five models.
we summarize the event rates with TianQin for all five models in the first column. 
In the second column we show the detection results for one TianQin set, and in the third column, we show the detection results for the twin constellations configuration. 
Notice that the rate for twin constellations is about double the one TianQin set, thanks to the relay in observation.
And for the heavy-seed model, most mergers occurring anywhere in the Universe can be detected with twin constellations. 

It is important to notice the 3 orders-of-magnitude difference among different models. 
Such a difference is partly due to the lack of resolution of the Millennium-I models and partly reflects the status quo of our current knowledge of galaxy and \ac{MBH} evolution. 
Although different models are calibrated against a number of observations, uncertainties in the early evolution of (proto)galactic structures and the \acp{MBH} hosted within them result in vastly different predictions of the \ac{MBH} binary merger and detection rates.

% compare with other results on LISA
For a sanity check, we compare our Table \ref{event_rate_all} with previous work. 
For example, the fully hydrodynamical cosmological simulation EAGLE  was also used to carry out a study of the \ac{GW} detection rate from \ac{MBH} binary mergers. 
Although the simulation method is different from the semianalytical model of GABE, the seeding mechanism and mass halo resolution are similar. 
As a result, Ref. \cite{EAGLE2016} obtained a detection rate of approximately two per year for eLISA, which is very close to our results for twin constellations when the H-seed model is considered. 
The results of the popIII, Q3\_d, and Q3\_nod models can be directly compared to the detection rates presented in Ref. \cite{Klein16} for LISA. The detection rates of LISA and twin constellations are very similar. 
This is because the merger rate is dominated by relatively low-mass systems, which fall in the sweet spot of both detectors. 

% They are consistent!
The reason for the much lower detection rate predicted by cosmological simulation-based models (both GABE and EAGLE) is at least partially subject to the limited mass resolution. 
Due to the huge computational cost, it is in fact currently infeasible to resolve low-mass halos within a large simulation box. 
Conversely, with the price of being more {\it ad hoc}, analytical EPS models are computationally cheaper and can reconstruct the halo merger history to much lower masses. 
We can therefore consider the  GABE/EAGLE results as an absolute lower boundary, and the popIII/Q3\_d/Q3\_nod as more fiducial estimates, although the Q3\_nod model is likely optimistic due to the absence of \ac{MBH} binary merger delays.

\begin{table}
    \begin{center}
%    \begin{adjustbox}{max width=\textwidth}
        \begin{tabular}{c|c|c|c|c|c}
            \hline
            \hline
            \multicolumn{1}{c|}{\multirow{2}{*}{Model}} &
            \multicolumn{1}{c|}{\multirow{2}{*}{Event rate($yr^{-1}$)}}&
            \multicolumn{2}{c|}{TianQin}&\multicolumn{2}{c}{Twin constellations}\\ \cline{3-6}
            \multicolumn{1}{c|}{}&\multicolumn{1}{c|}{}&
            Detection rate($yr^{-1}$)&Detection percentage&Detection rate($yr^{-1}$)&Detection percentage \\
            \hline
            $\rm{L-seed}$&2.57&0.08&3.1\%&0.162&6.3\%\\
            \hline
            $\rm{H-seed}$&2.57&1.055&41.1\%&1.642&63.9\%\\
            \hline
            popIII&174.70&10.58&6.1\%&22.60&12.9\%\\
            \hline
            $\rm{Q3\_d}$&8.18&4.42&54.0\%&8.06&98.5\%\\
            \hline
            $\rm{Q3\_nod}$&122.44&58.96&48.2\%&118.12&96.5\%\\
            \hline
            \hline

        \end{tabular}
%    \end{adjustbox}
    \end{center}
    \caption{\label{event_rate_all} \ac{MBH} binary cosmic merger rates and TianQin detection rates for the five investigated \ac{MBH} population models. Detection rates are given considering both one and two TianQin detectors. We also show the percentage of the detection rate as a percentage of the event rate.}
\end{table}

%%%%%%%%%%%%%%%%%%%%%%%%%%%%%%%%%%%%%%%%%%%%%%%%%%%%%%%%%%%%%%%%
\subsection{Parameter estimation}

We perform \ac{FIM} analysis to determine the parameter estimation precision of TianQin detections. 
We consider two fiducial cases for demonstration.
First, if a \ac{MBH} binary merger happens shortly after formation, accurate determination of component black hole masses and luminosity distance could help to distinguish seed models.
Second, for nearby \ac{MBH} mergers, an early warning before merger, with a forecast of the sky location as well as merger time, could be greatly helpful for the preparation of multimessenger observations.

In the calculation of the two fiducial cases, we set $\chi_1=0,\,\chi_2=0$, $\theta=\pi/3$, $\phi=0$, $\iota=\pi/2$, $\psi=\pi/3$, $\phi_c=-\pi/4$, and $t_c=3$ months, where the value of $\theta$ is chosen to be representitive without loss of generality. 
Notice that we retain the geometric factor of $\sqrt{3}/2$ in the inner product.

\begin{figure}[htb]
\centering
  \begin{subfigure}[b]{.30\linewidth}
    \centering
    \includegraphics[width=.99\textwidth,height=5cm]{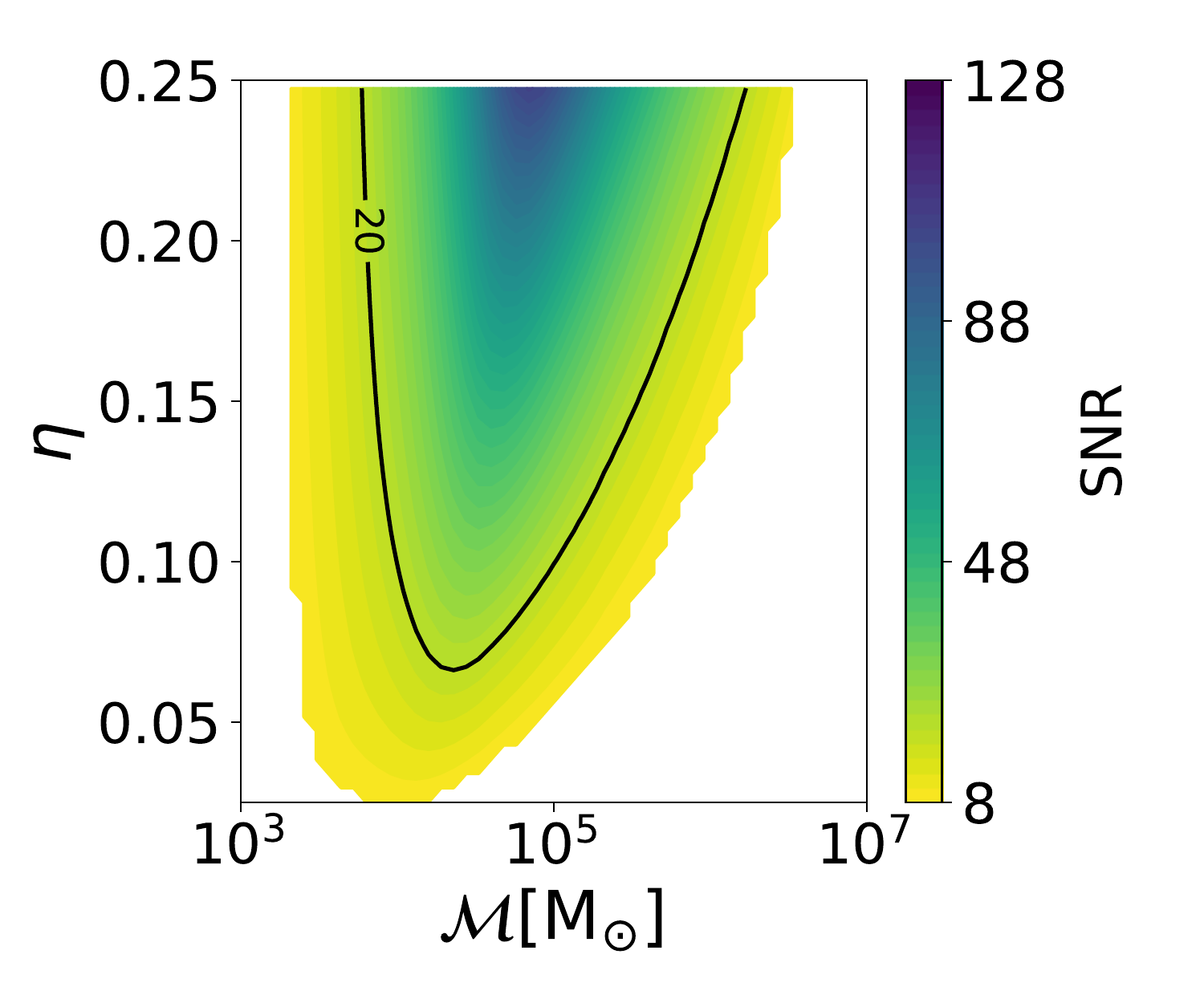}
    \caption{Relative SNR.}\label{snr1_m35}
  \end{subfigure}%
  \begin{subfigure}[b]{.30\linewidth}
    \centering
    \includegraphics[width=.99\textwidth,height=5cm]{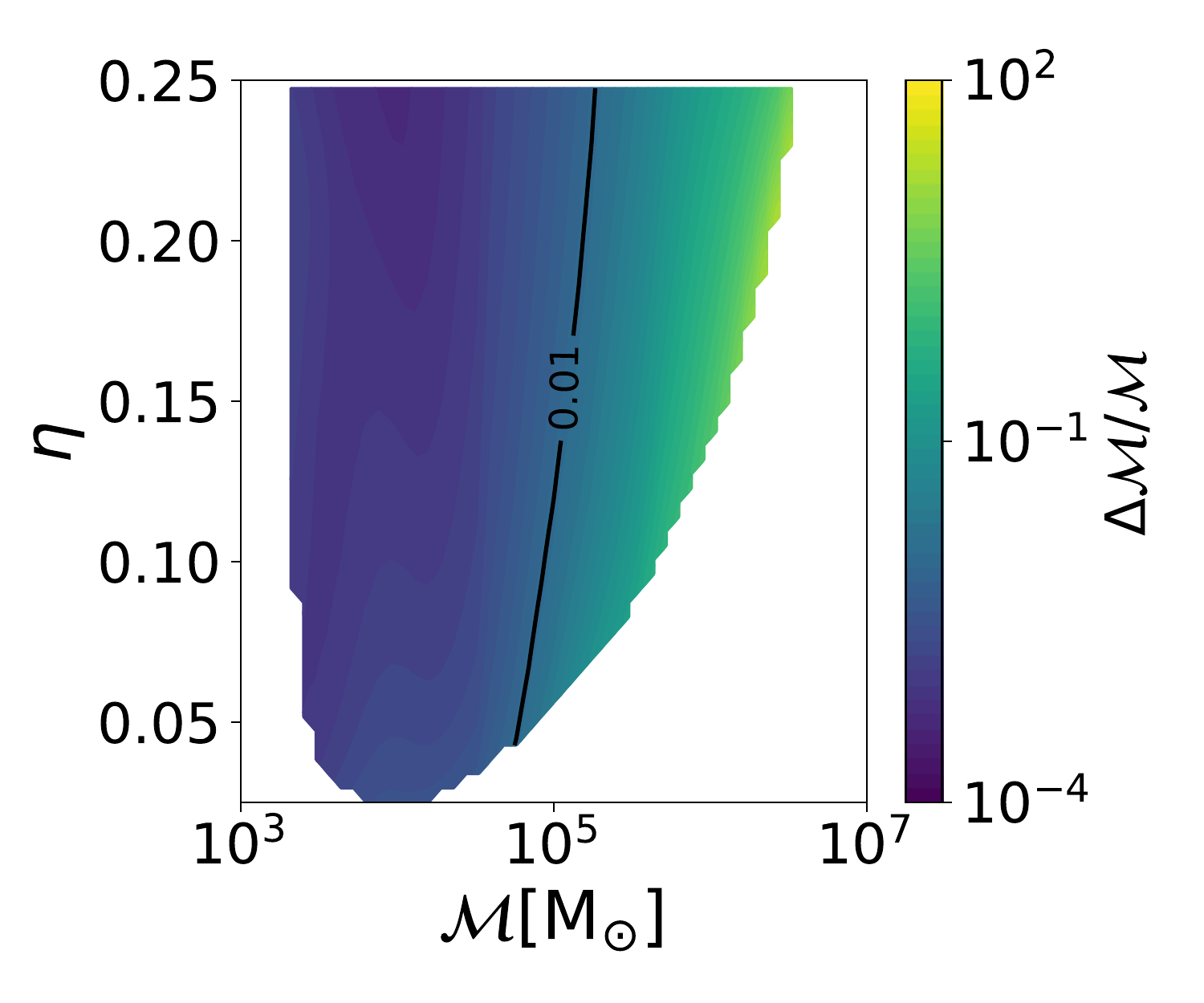}
    \caption{Relative uncertainty of $\mathcal{M}$.}\label{cm_m35}
  \end{subfigure}%
  \begin{subfigure}[b]{.30\linewidth}
    \centering
    \includegraphics[width=.99\textwidth,height=5cm]{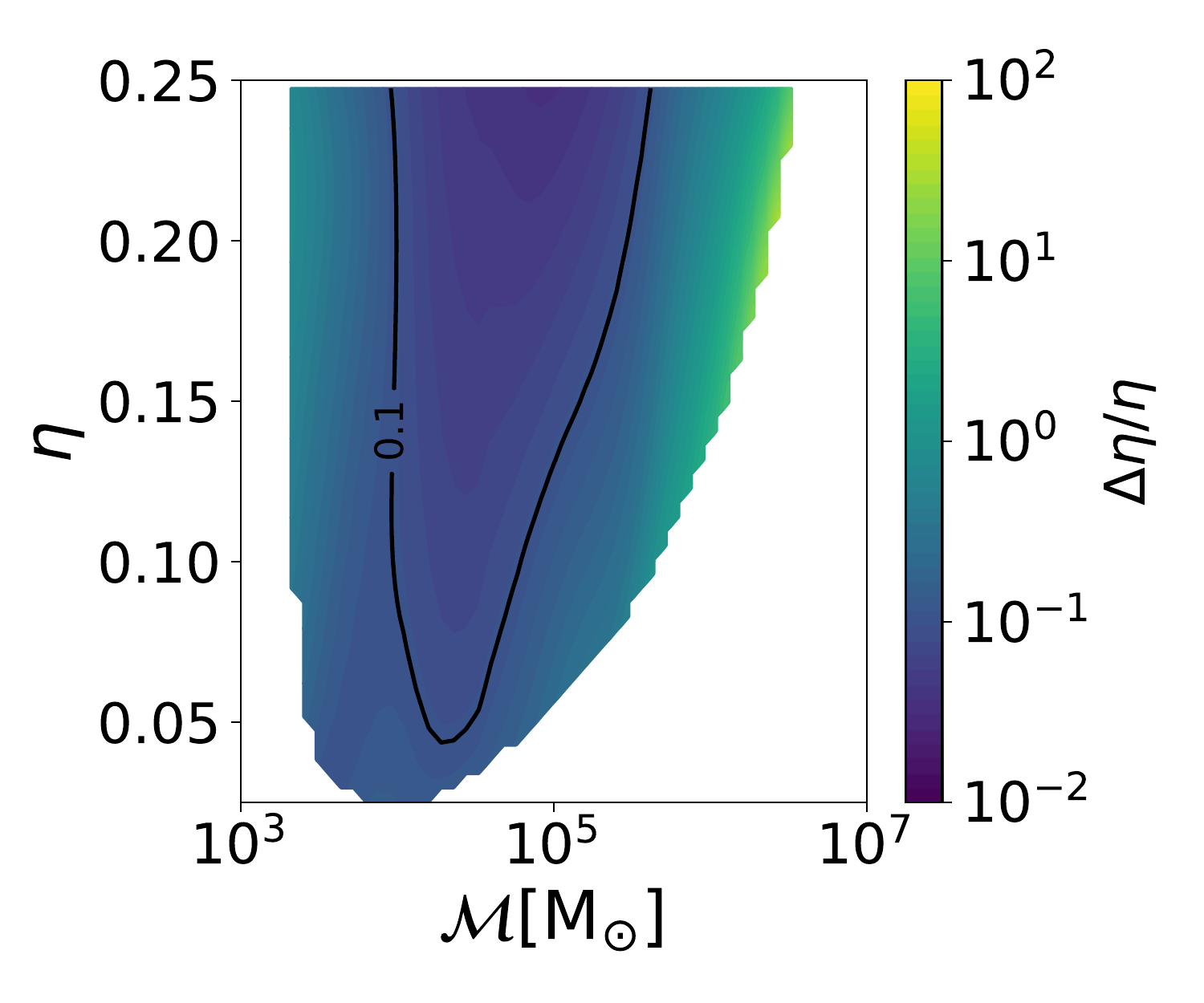}
    \caption{Relative uncertainty of $\eta$.}\label{eta_m35}
  \end{subfigure}\\%
  \begin{subfigure}[b]{.30\linewidth}
    \centering
    \includegraphics[width=.99\textwidth,height=5cm]{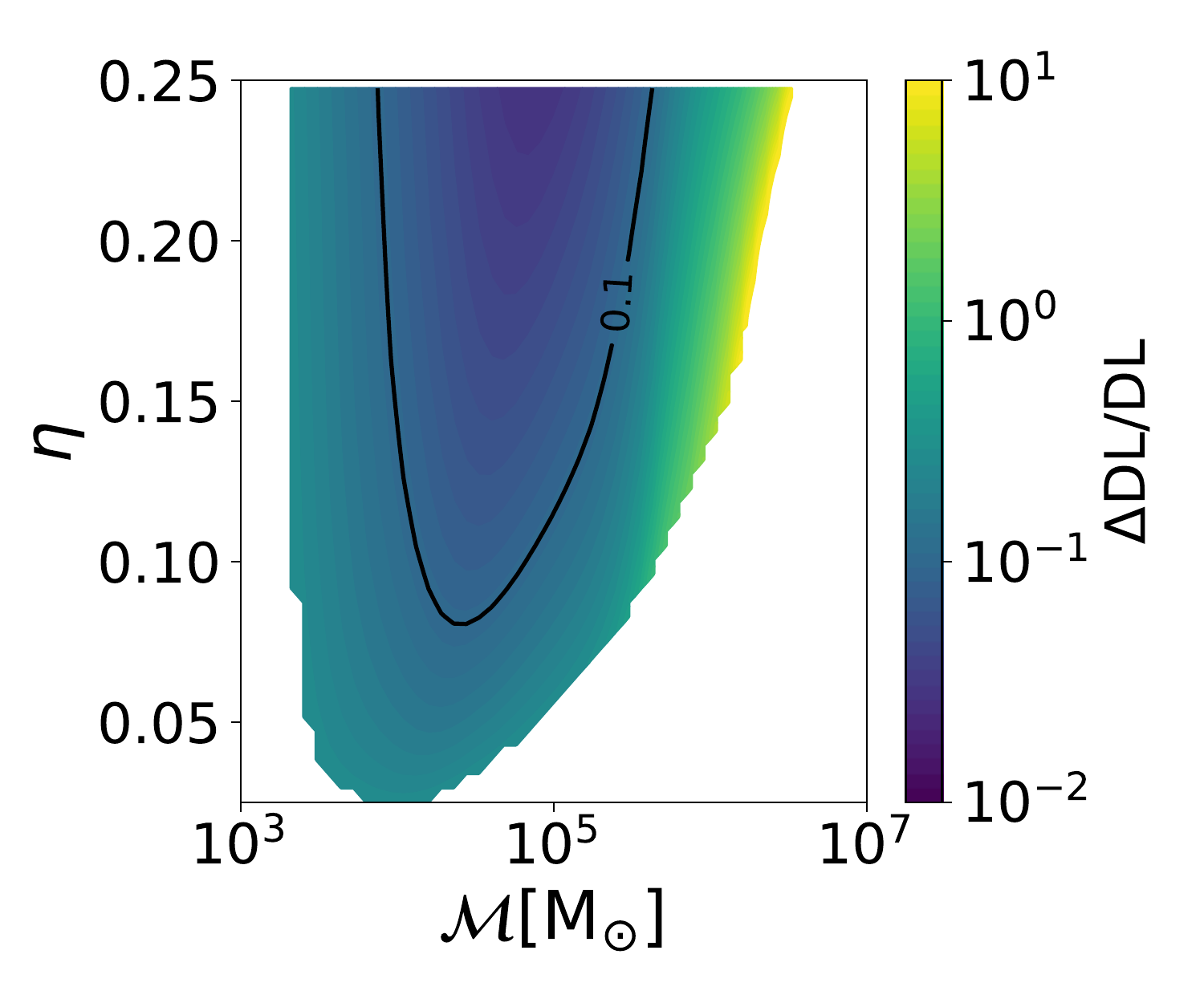}
    \caption{Relative uncertainty of $D_L$.}\label{dl_m35}
  \end{subfigure}%
  \begin{subfigure}[b]{.30\linewidth}
    \centering
    \includegraphics[width=.99\textwidth,height=5cm]{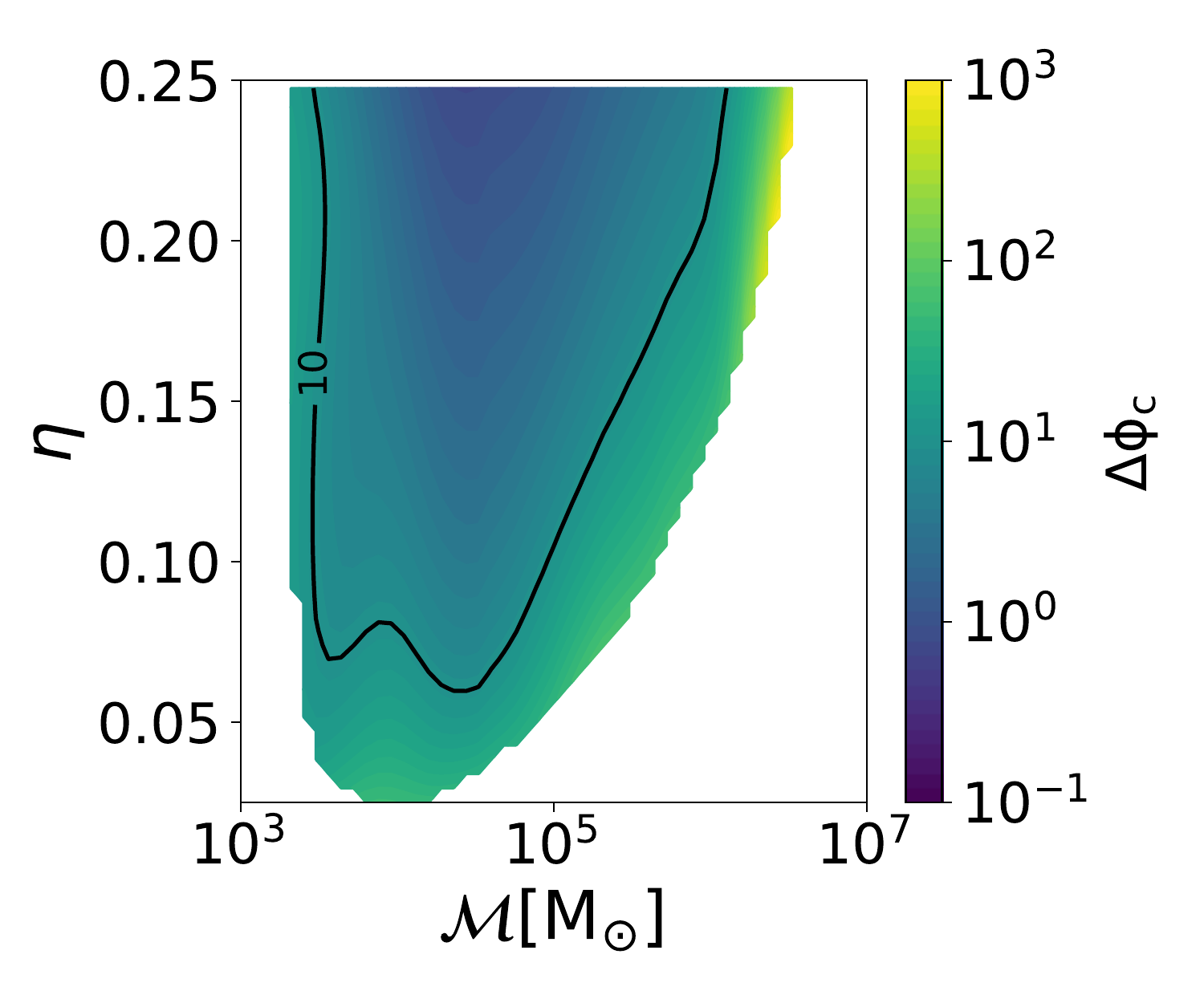}
    \caption{Relative uncertainty of $\phi_c$.}\label{phic_m35}
  \end{subfigure}%
  \begin{subfigure}[b]{.30\linewidth}
    \centering
    \includegraphics[width=.99\textwidth,height=5cm]{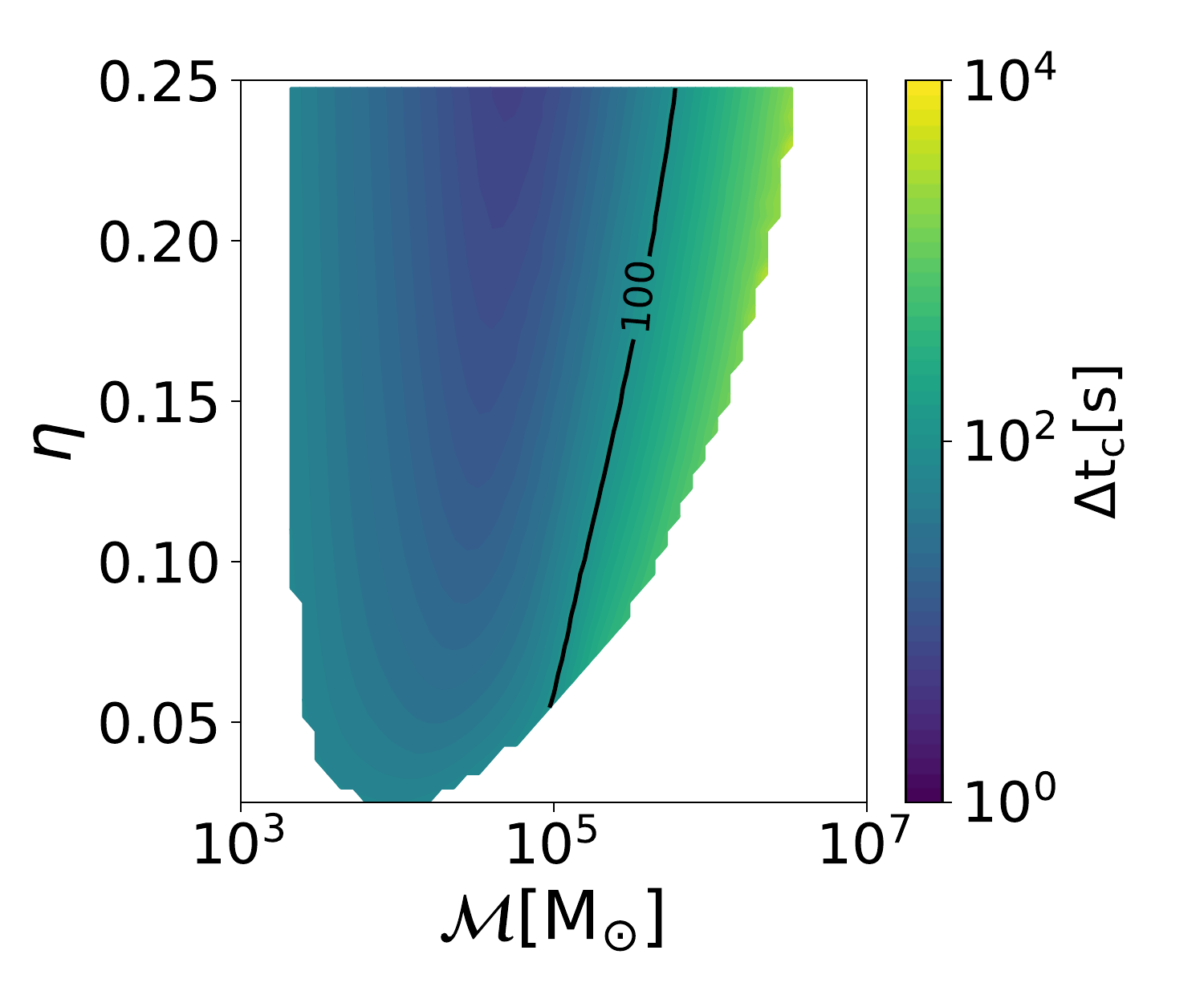}
    \caption{Relative uncertainty of $t_c$.}\label{tc_m35}
  \end{subfigure}\\%
  \begin{subfigure}[b]{.30\linewidth}
    \centering
    \includegraphics[width=.99\textwidth,height=5cm]{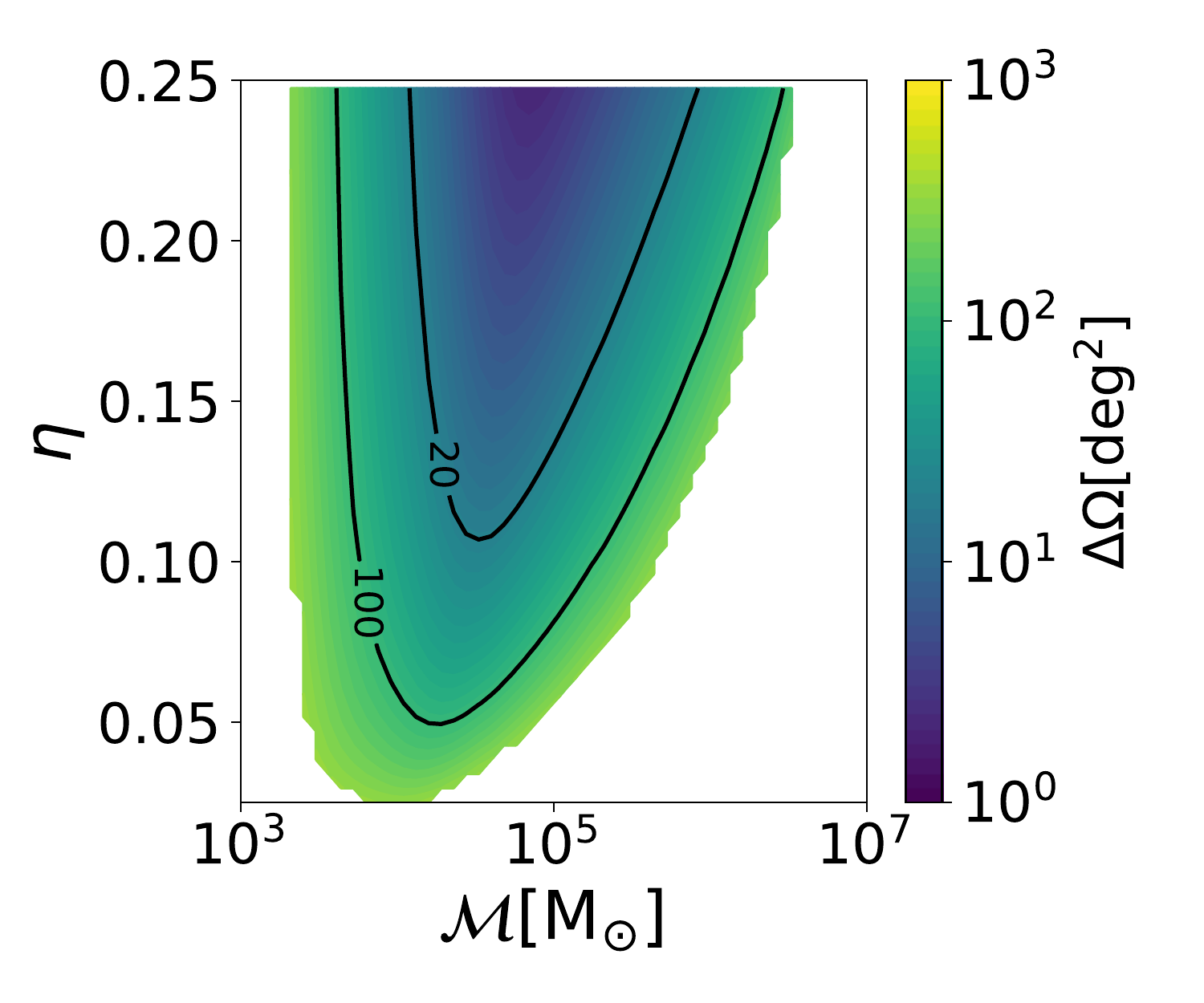}
    \caption{Relative uncertainty of $\Omega$.}\label{loc_m35}
  \end{subfigure}%
  \begin{subfigure}[b]{.30\linewidth}
    \centering
    \includegraphics[width=.99\textwidth,height=5cm]{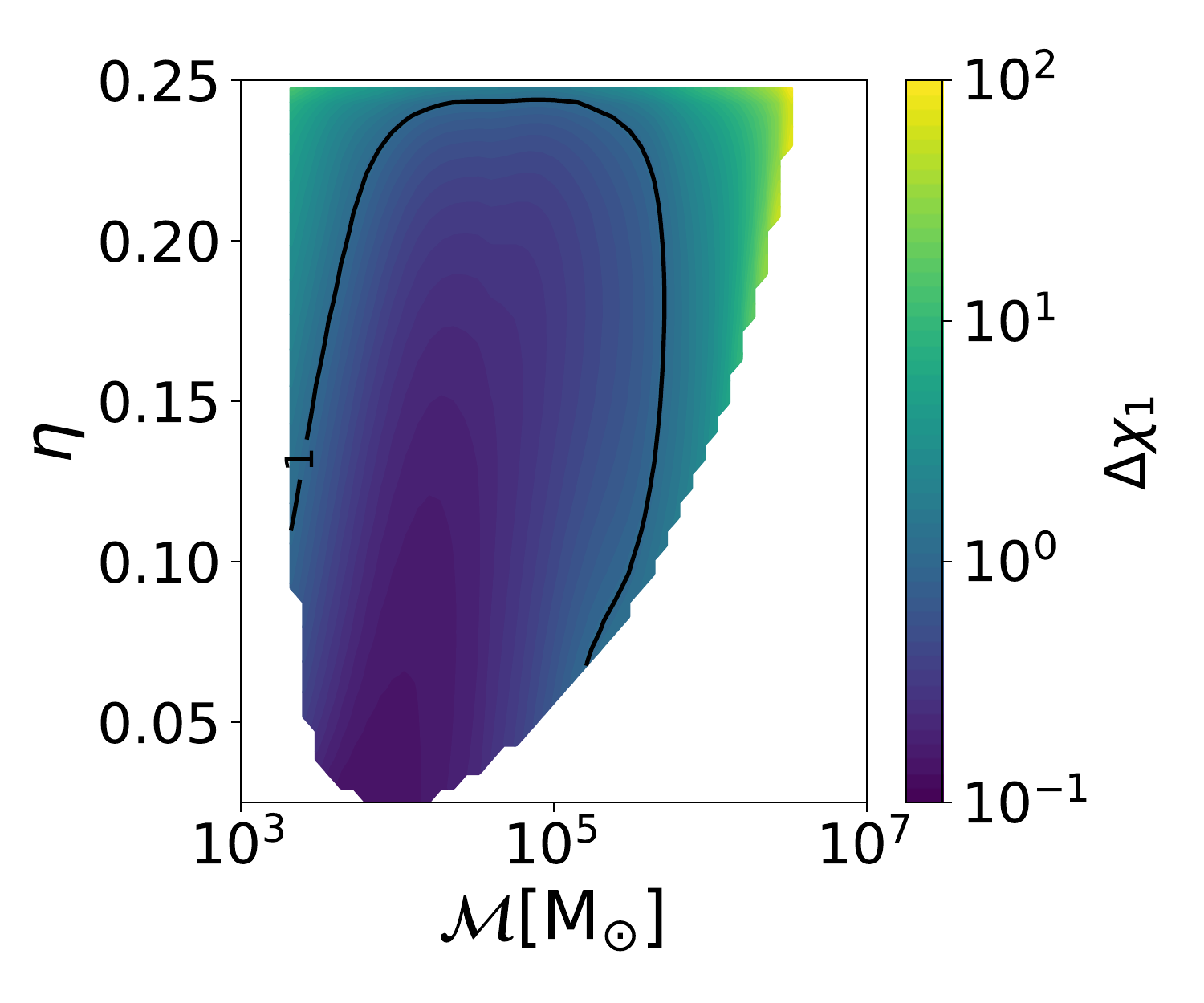}
    \caption{Relative uncertainty of $\chi_1$.}\label{chi1_m35}
  \end{subfigure}%
  \begin{subfigure}[b]{.30\linewidth}
    \centering
    \includegraphics[width=.99\textwidth,height=5cm]{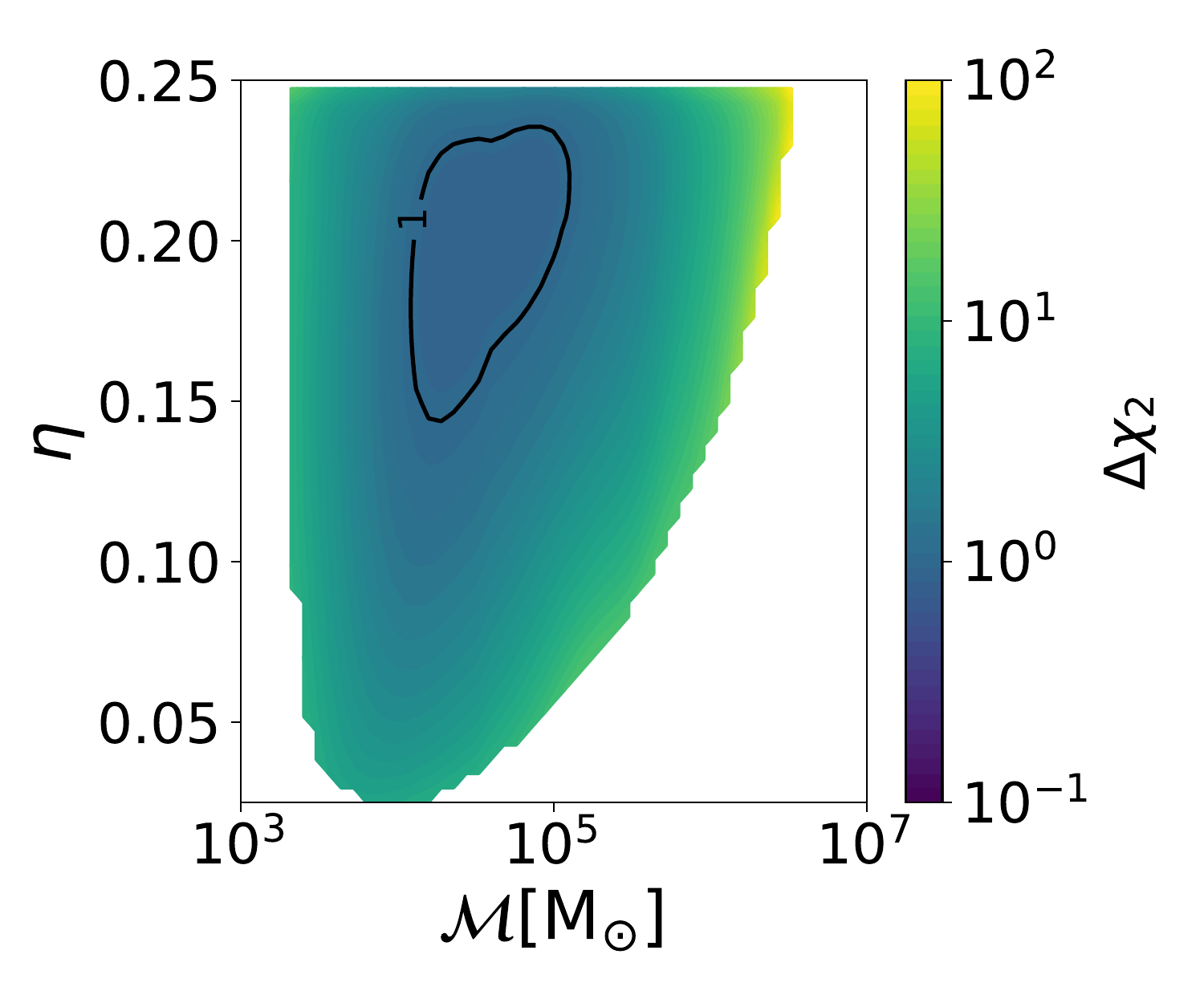}
    \caption{Relative uncertainty of $\chi_2$.}\label{chi2_m35}
  \end{subfigure}\\%
\caption{The contour of relative parameter estimation error on different parameters, assuming a redshift $z=15$.
%chirp mass (left panel) and on luminosity distance (right panel) at $z=15$.
Signals are assumed to last for three months before they merge in the TianQin band.
Only events with SNR $\geq 8$ are shown. 
We mark certain fiducial values with black lines. }
\label{fig:m3m5z15}
\end{figure}

% D_l uncertainty (seed model)

For the first scenario, we consider a merger which happened at redshift $z=15$, and in Fig. \ref{fig:m3m5z15}, we present the distribution of SNR as well as expected uncertainties of parameters over the parameter space of chirp mass $\mathcal{M}$ and symmetric mass-ratio $\eta$.
Figure. \ref{snr1_m35} indicates that at a redshift of 15, the optimal SNR can be as high as $120$.
We notice from Fig. \ref{cm_m35} and \ref{dl_m35} and that if both \acp{MBH} are in the range $10^4\,M_{\odot}<\mathcal{M}<10^6\,M_{\odot}$, TianQin can reach a fractional error of $10\%$ for luminosity distance, and the fractional error of the chirp mass can be as high as $10\%$ for sources with SNR around $20$. 
Therefore, if a \ac{MBH} binary with masses around $10^4-10^5 M_\odot$ merges at high redshift, its masses and luminosity distance can be estimated with sufficient accuracy, so it is possible to distinguish different seed models.

For the symmetric mass-ratio determination $\Delta \eta/{\eta}$, TianQin can reach a fractional error of $10\%$ for symmetric mass-ratio with chirp mass in the range $10^4\,M_{\odot}<\mathcal{M}<10^5\,M_{\odot}$ and symmetric mass-ratio higher than 0.05.
For most of the sources with chirp mass in the range $10^3\Msun<{\cal M}<10^7\Msun$, TianQin can make a detection with a sky location error of less than $100\ \text{deg}^2$. When the source frame chirp mass is in the range $10^4\,M_{\odot}<\mathcal{M}<10^6\,M_{\odot}$ and the symmetric mass-ratio is higher than 0.1, the sky location error can be better than $20$ $\text{deg}^2$.
The error of $\phi_c$ is more sensitive to the symmetric mass-ratio than the chirp mass, which is different from other results in Fig. \ref{fig:m3m5z15}. 
In terms of timing ability, TianQin can constrain the merger time $\rm t_c$ with accuracy better than $100$s. 
Notice that the uncertainty on spin $\chi_1$ is constantly better than $\chi_2$. 
This is due to the fact that the uncertainty on $\chi_{1,2}$ is strongly dependent on the value of mass $\rm m_{1,2}$, while we set $m_1\geq m_2$; thus, the aforementioned observation of the better constraint on $\chi_{1}$ is naturally expected.

\begin{figure}[htb]
\centering
  \begin{subfigure}[b]{.30\linewidth}
    \centering
    \includegraphics[width=.99\textwidth,height=5cm]{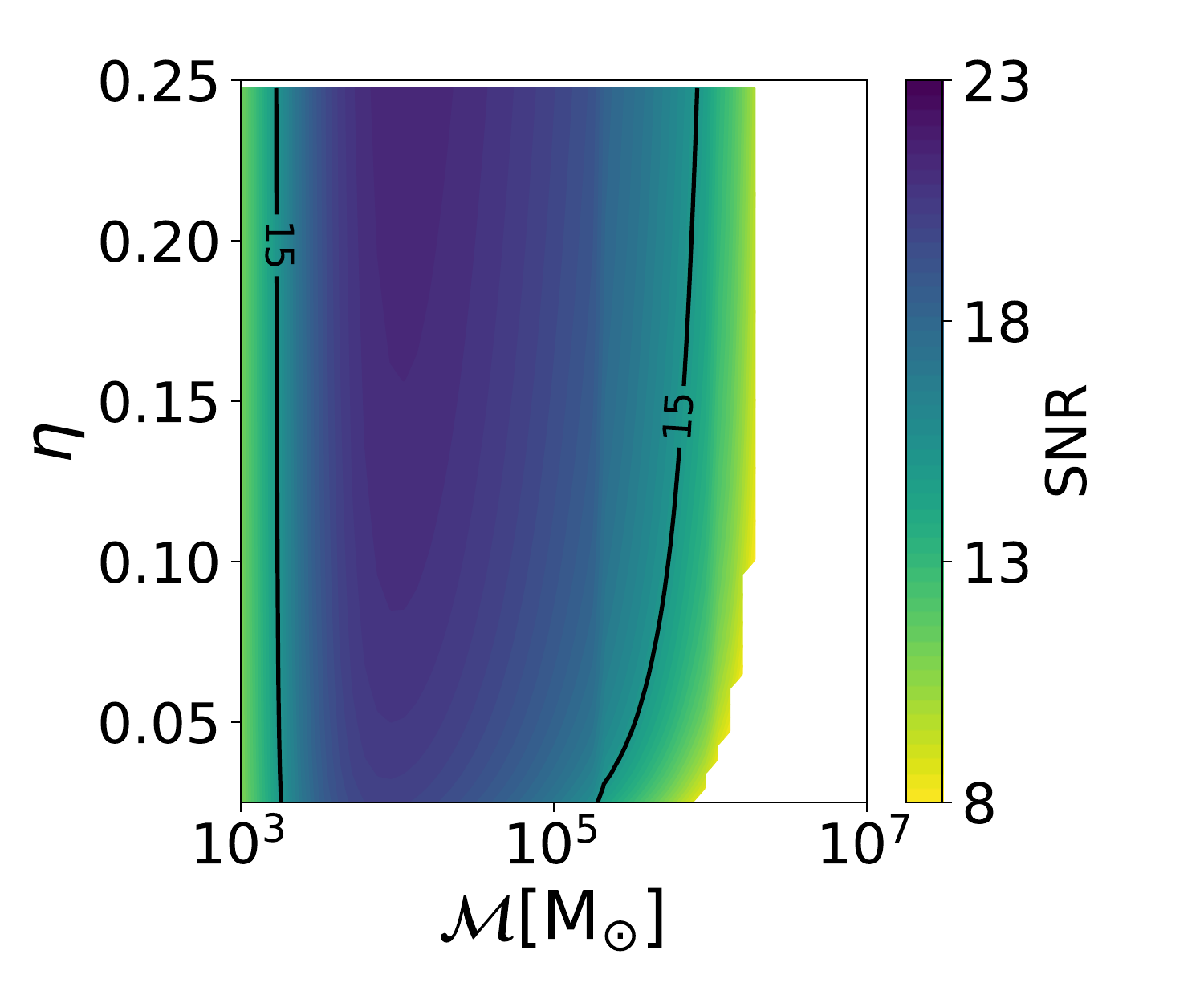}
    \caption{Relative SNR.}\label{snr1_m47}
  \end{subfigure}%
  \begin{subfigure}[b]{.30\linewidth}
    \centering
    \includegraphics[width=.99\textwidth,height=5cm]{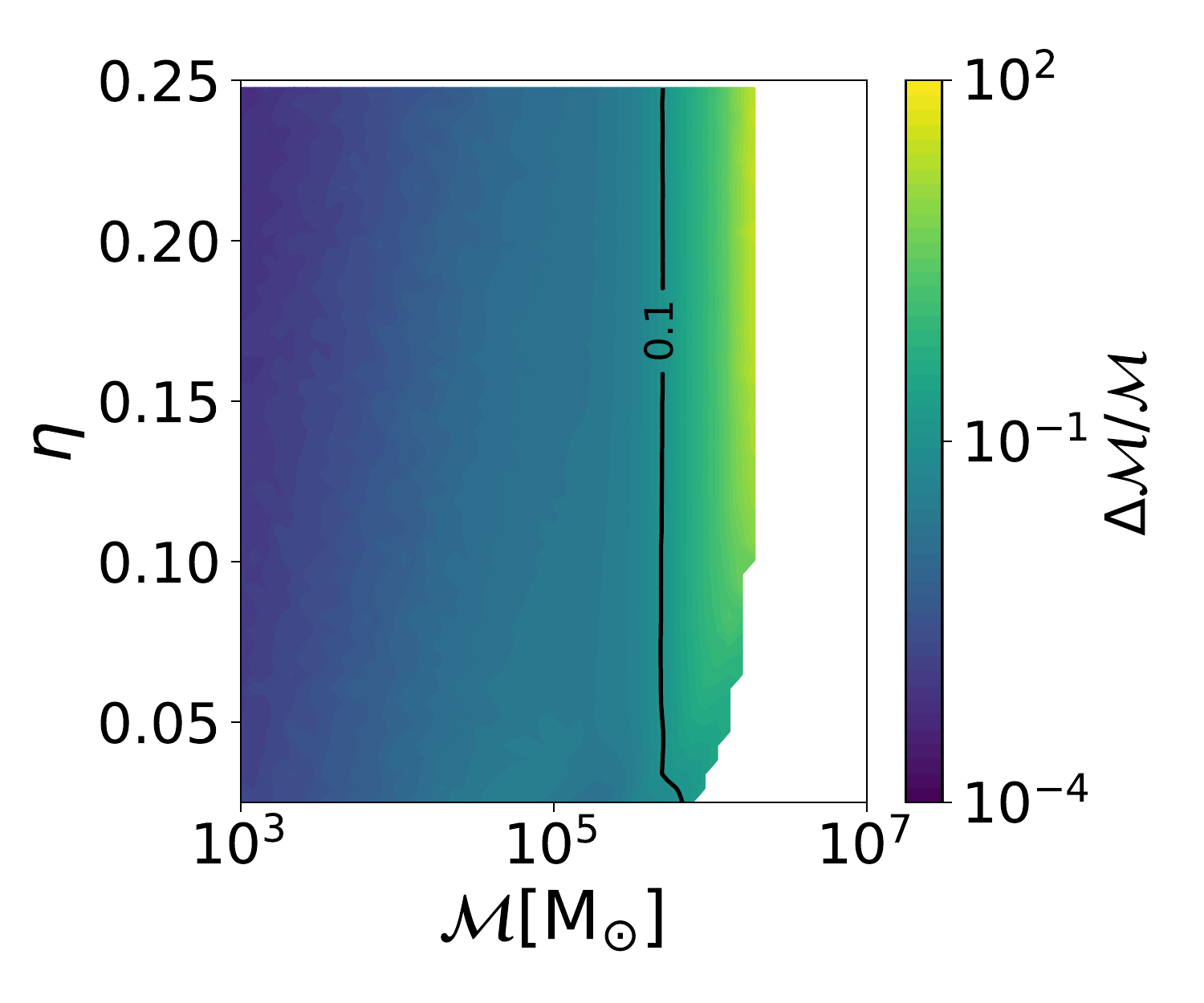}
    \caption{Relative uncertainty of $\mathcal{M}$.}\label{cm_m47}
  \end{subfigure}%
  \begin{subfigure}[b]{.30\linewidth}
    \centering
    \includegraphics[width=.99\textwidth,height=5cm]{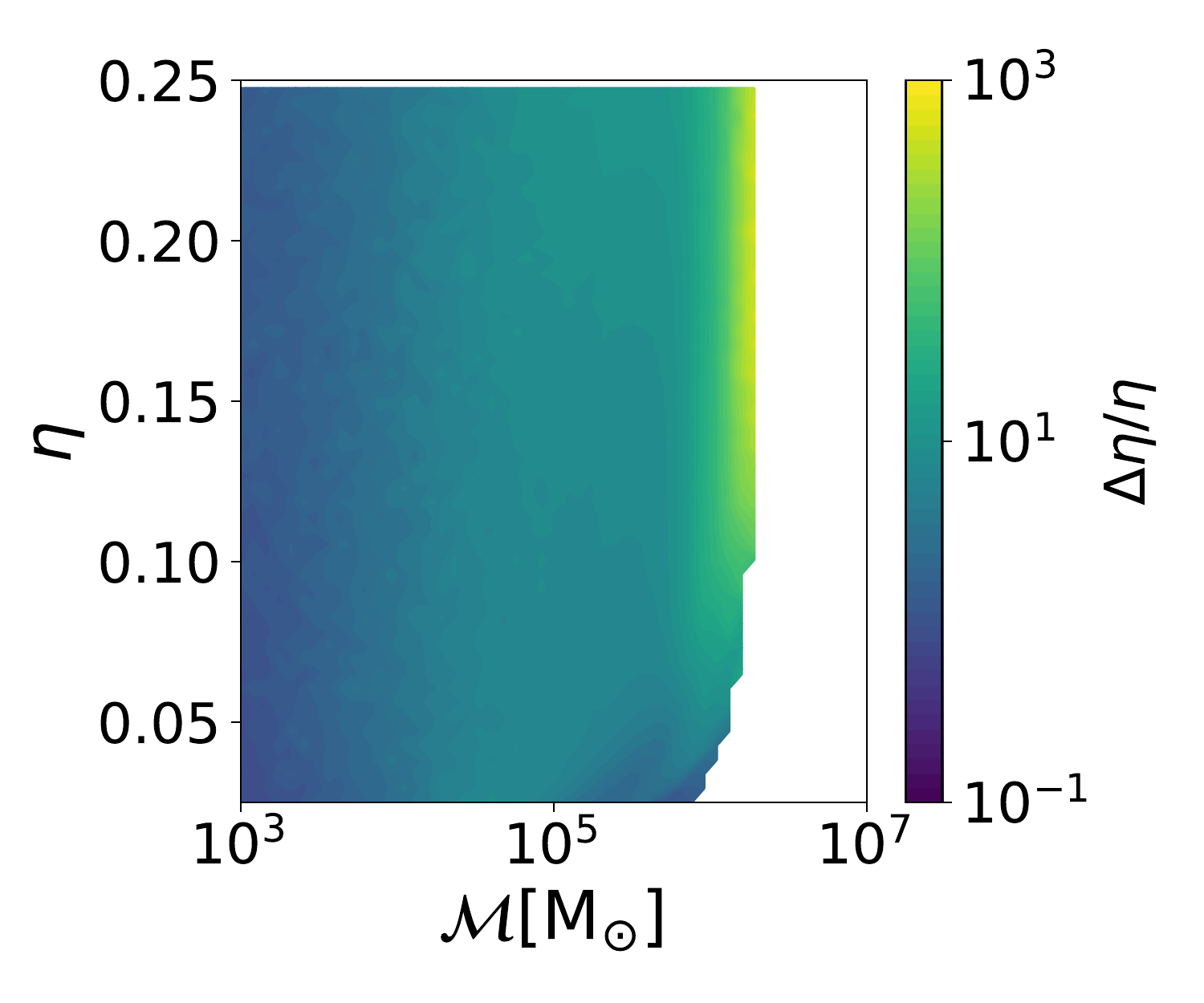}
    \caption{Relative uncertainty of $\eta$.}\label{eta_m47}
  \end{subfigure}\\%
  \begin{subfigure}[b]{.30\linewidth}
    \centering
    \includegraphics[width=.99\textwidth,height=5cm]{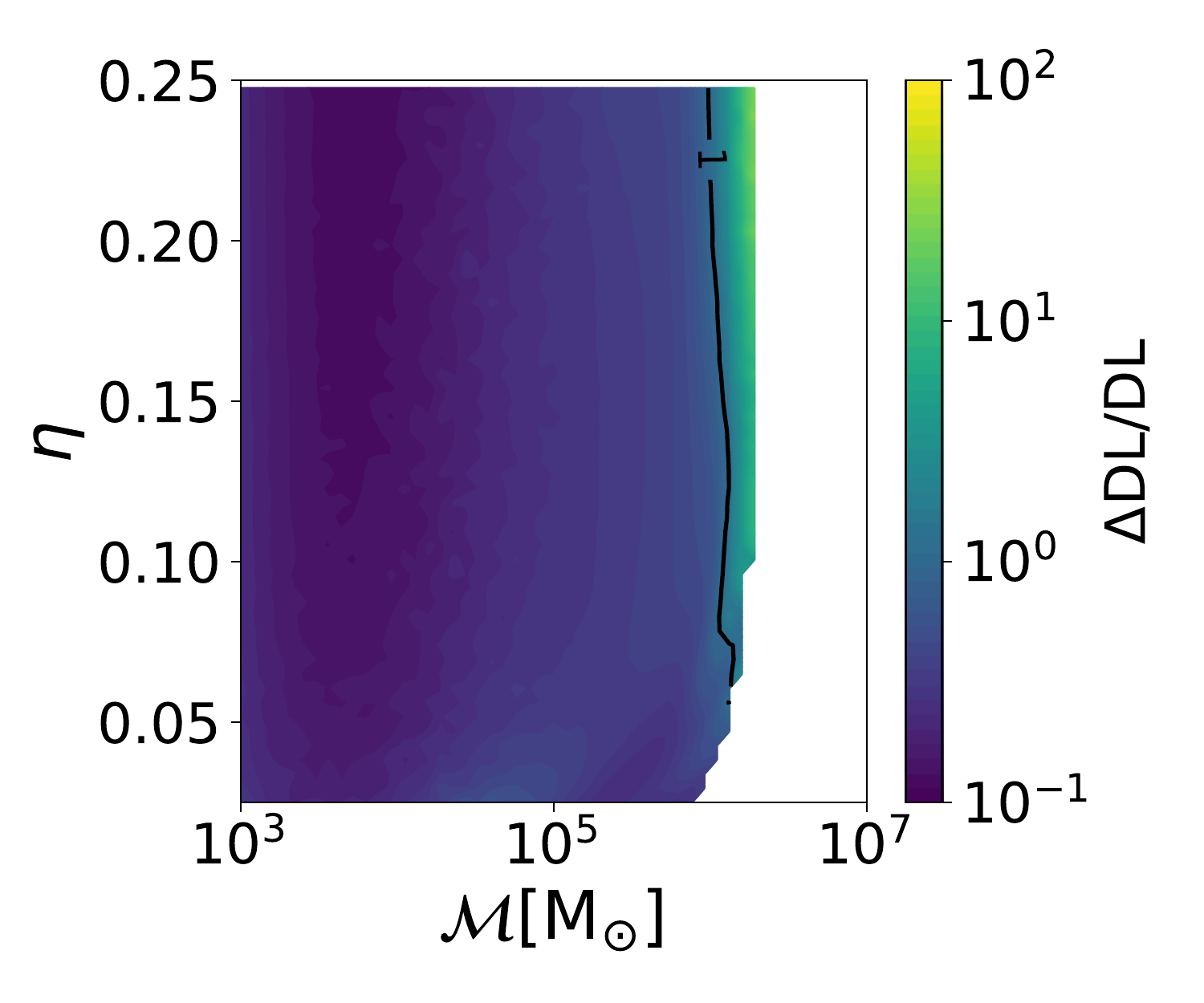}
    \caption{Relative uncertainty of $D_L$.}\label{dl_m47}
  \end{subfigure}%
  \begin{subfigure}[b]{.30\linewidth}
    \centering
    \includegraphics[width=.99\textwidth,height=5cm]{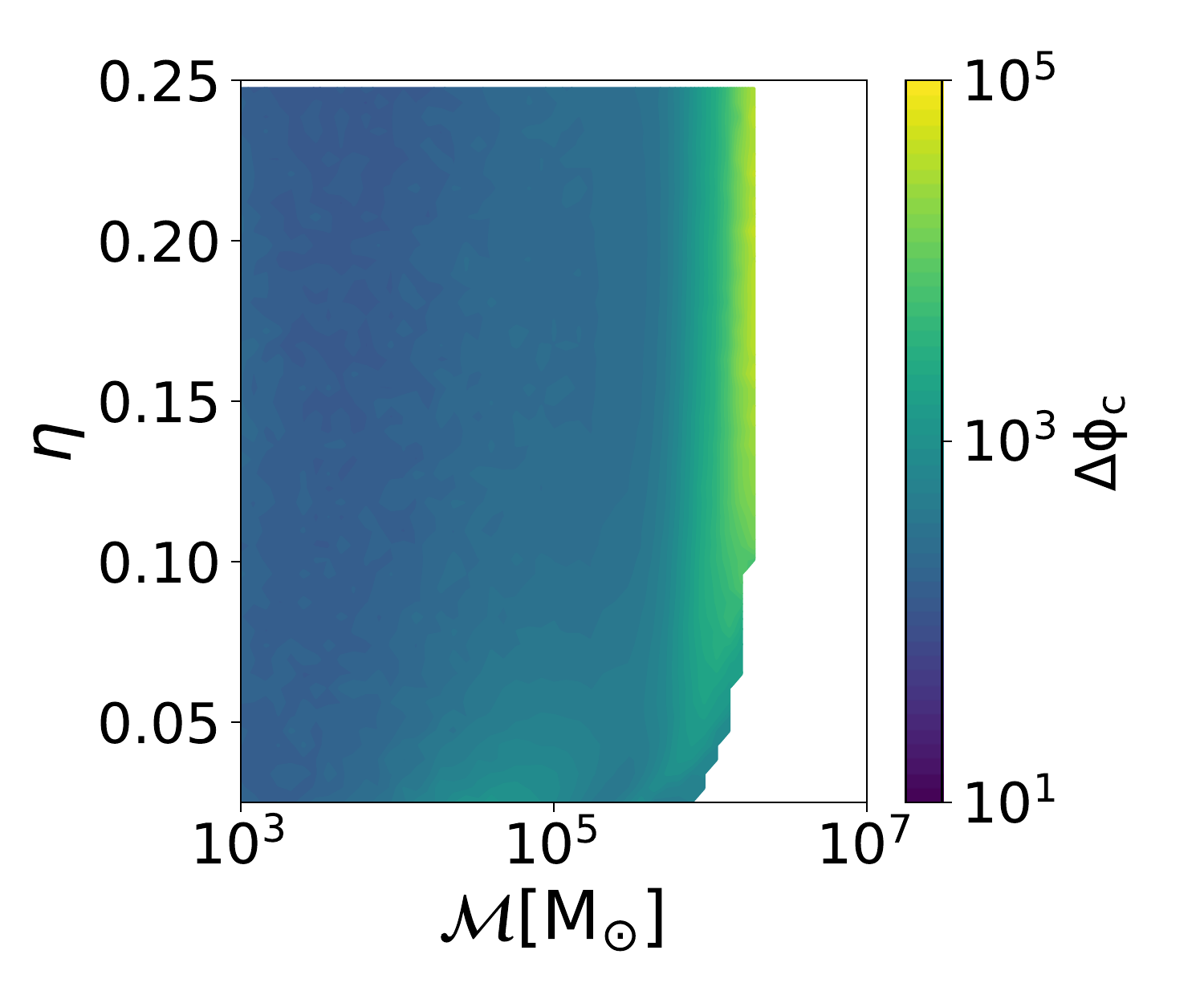}
    \caption{Relative uncertainty of $\phi_c$.}\label{phic_m47}
  \end{subfigure}%
  \begin{subfigure}[b]{.30\linewidth}
    \centering
    \includegraphics[width=.99\textwidth,height=5cm]{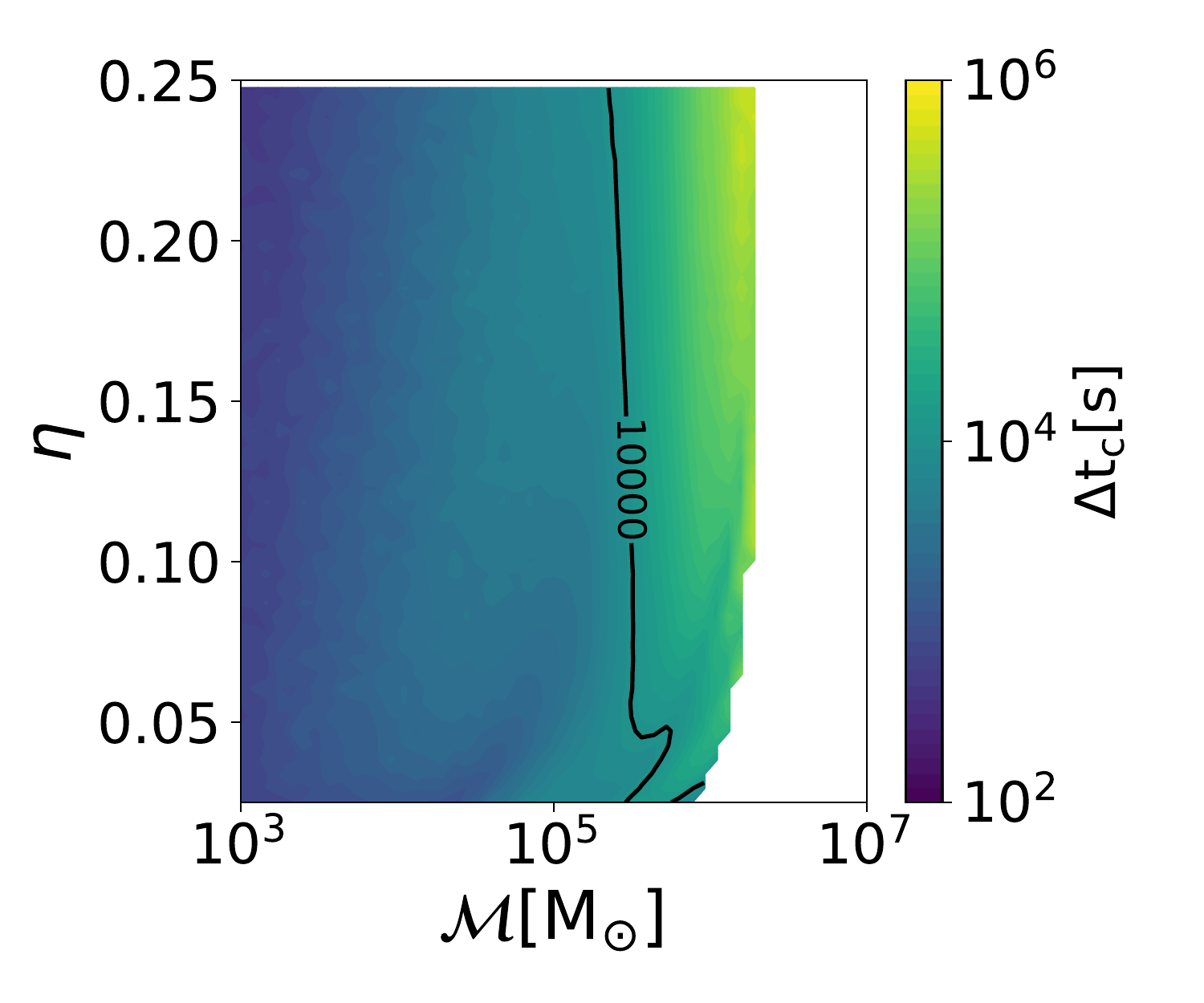}
    \caption{Relative uncertainty of $t_c$.}\label{tc_m47}
  \end{subfigure}\\%
  \begin{subfigure}[b]{.30\linewidth}
    \centering
    \includegraphics[width=.99\textwidth,height=5cm]{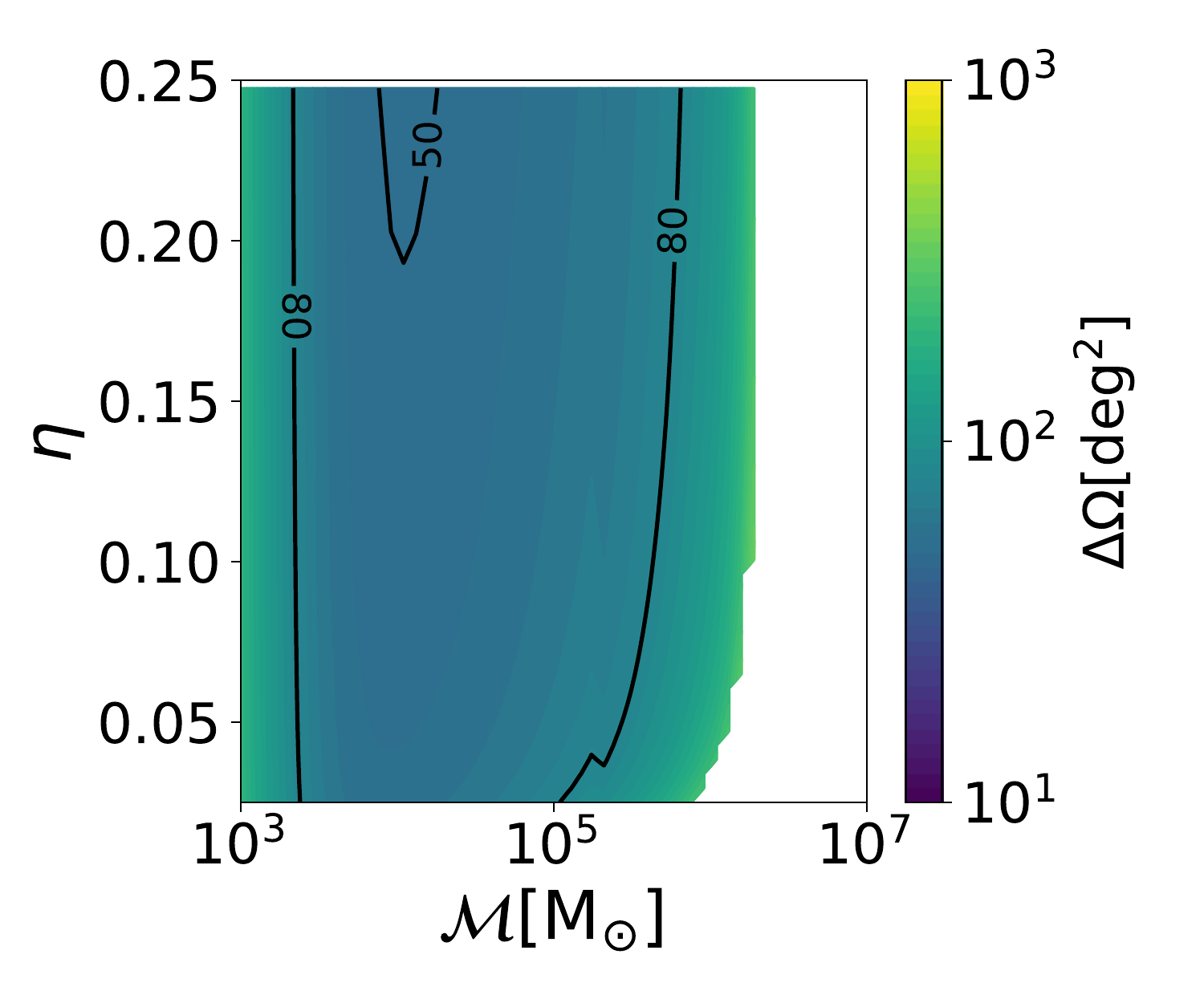}
    \caption{Relative uncertainty of $\Omega$.}\label{loc_m47}
  \end{subfigure}%
  \begin{subfigure}[b]{.30\linewidth}
    \centering
    \includegraphics[width=.99\textwidth,height=5cm]{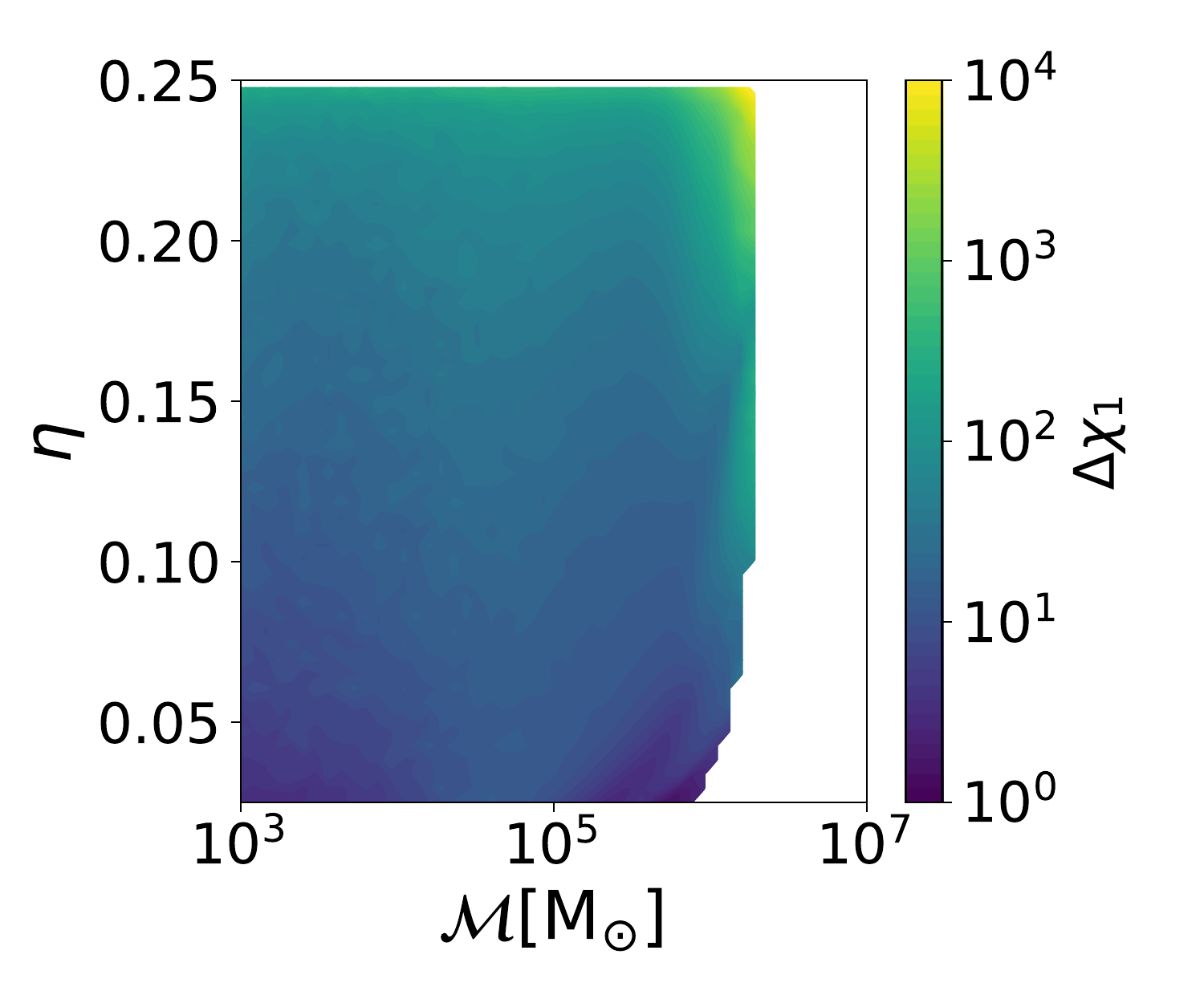}
    \caption{Relative uncertainty of $\chi_1$.}\label{chi1_m47}
  \end{subfigure}%
  \begin{subfigure}[b]{.30\linewidth}
    \centering
    \includegraphics[width=.99\textwidth,height=5cm]{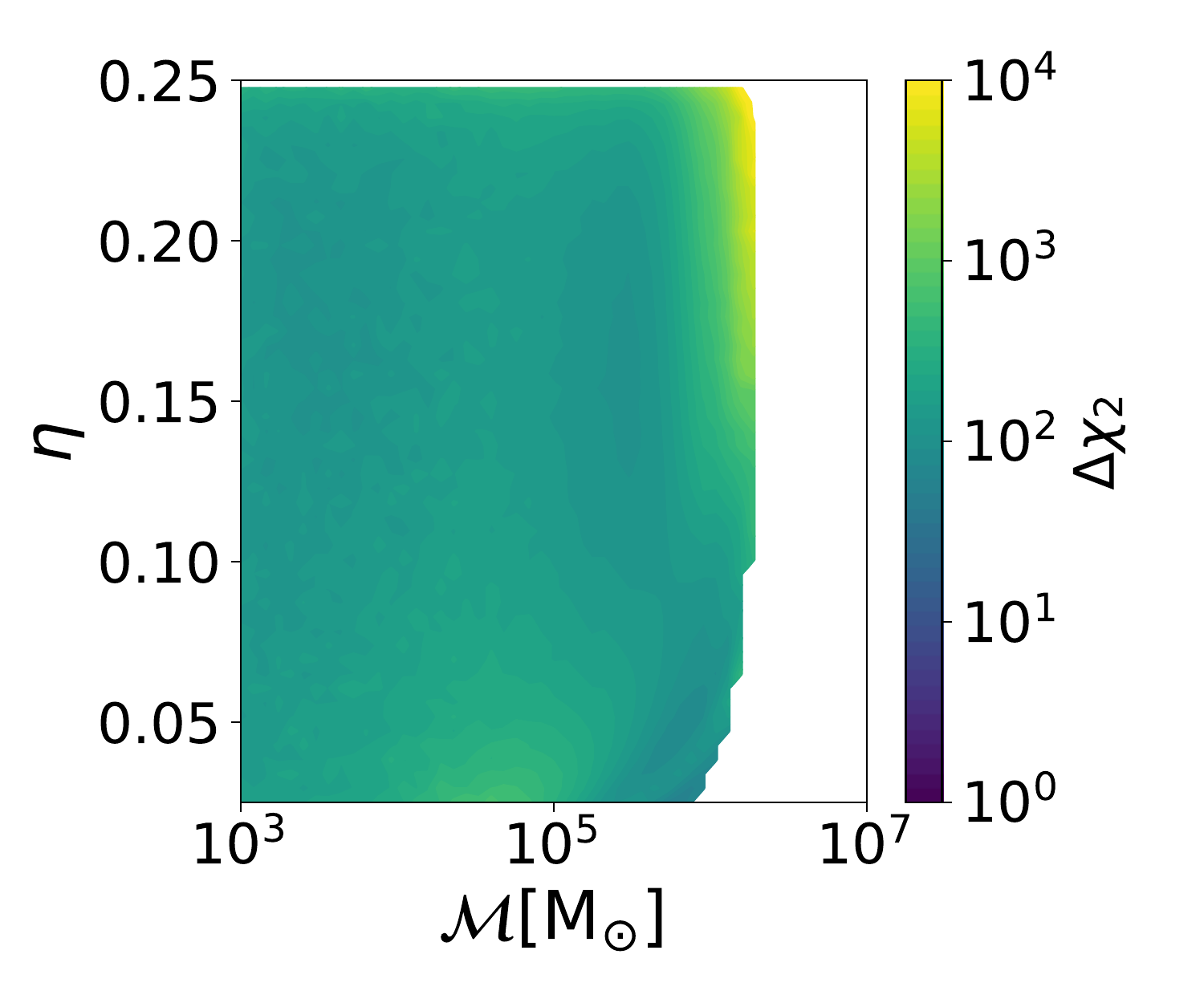}
    \caption{Relative uncertainty of $\chi_2$.}\label{chi2_m47}
  \end{subfigure}\\%
\caption{Early warning ability of TianQin for sources at $z=2$ as a function of chirp mass and symmetric mass-ratio. Contours of $\Delta{\Omega}$ (in deg$^2$) are represented by black lines while the color scale encodes the SNR of the sources. Quantities are measured by integrating the signal up to 24 hours before final coalescence. The small fluctuation showed on some of the figures are due to the numerical instability in the process of matrix inversion.
}
\label{fig:m4m7z2}
\end{figure}

% sky localisation
%The GABE model predicted, the  merger rate of MBHBs peaks at $z\sim 2$.
Although low-mass events at high redshift can be easily detected through the \ac{GW} channel, any putative electromagnetic counterpart will be extremely faint and likely beyond foreseeable observational capabilities. 
Counterpart identification will favor a lower redshift to lose the luminosity prerequisite of any potential radiation mechanisms.
We therefore explore sky localization performances for a fiducial redshift of $z = 2$, and we estimate the early warning ability of TianQin, focusing on the sky localization error $24$ hours before the merger. 
This would provide sufficient time to issue early warnings and point electromagnetic probes before the final coalescence.
%This would allow observing astronomer to have precious time to prepare for the EM observation of the merger.

We consider here a fiducial merger of \ac{MBH} at redshift of $z=2$.
In Fig. \ref{fig:m4m7z2}, we show distributions of parameter uncertainties over chirp mass $\mathcal{M}$ and symmetric mass-ratio $\eta$. 
For the majority of the sources with chirp mass in the range $10^4\Msun<{\cal M}<10^6\Msun$, TianQin can make a detection 24 hours before the final merger.
Notice that for sources at this redshift, integrating the signal until 24 hours before merger yields a maximum SNR of $23$. 
This enables us to issue early warning ahead of the actual merger, with a sky localization error of less than $100\ \text{deg}^2$ and a timing error of  round three hours or smaller. 
For optimal events, the sky localization could be better than $50\ {\rm deg}^2$, while the fractional error in the chirp mass estimate can be as high as $10\%$ when the SNR is around $15$ or higher. 
This sky localization accuracy is sufficiently small to be covered by finite number of exposures of future wide field of view instrument such as, for example, the large synoptic survey telescope \cite{2009arXiv0912.0201L}. 

\begin{figure}[htb]
\centering
  \begin{subfigure}[b]{.30\linewidth}
    \centering
    \includegraphics[width=.99\textwidth,height=5cm]{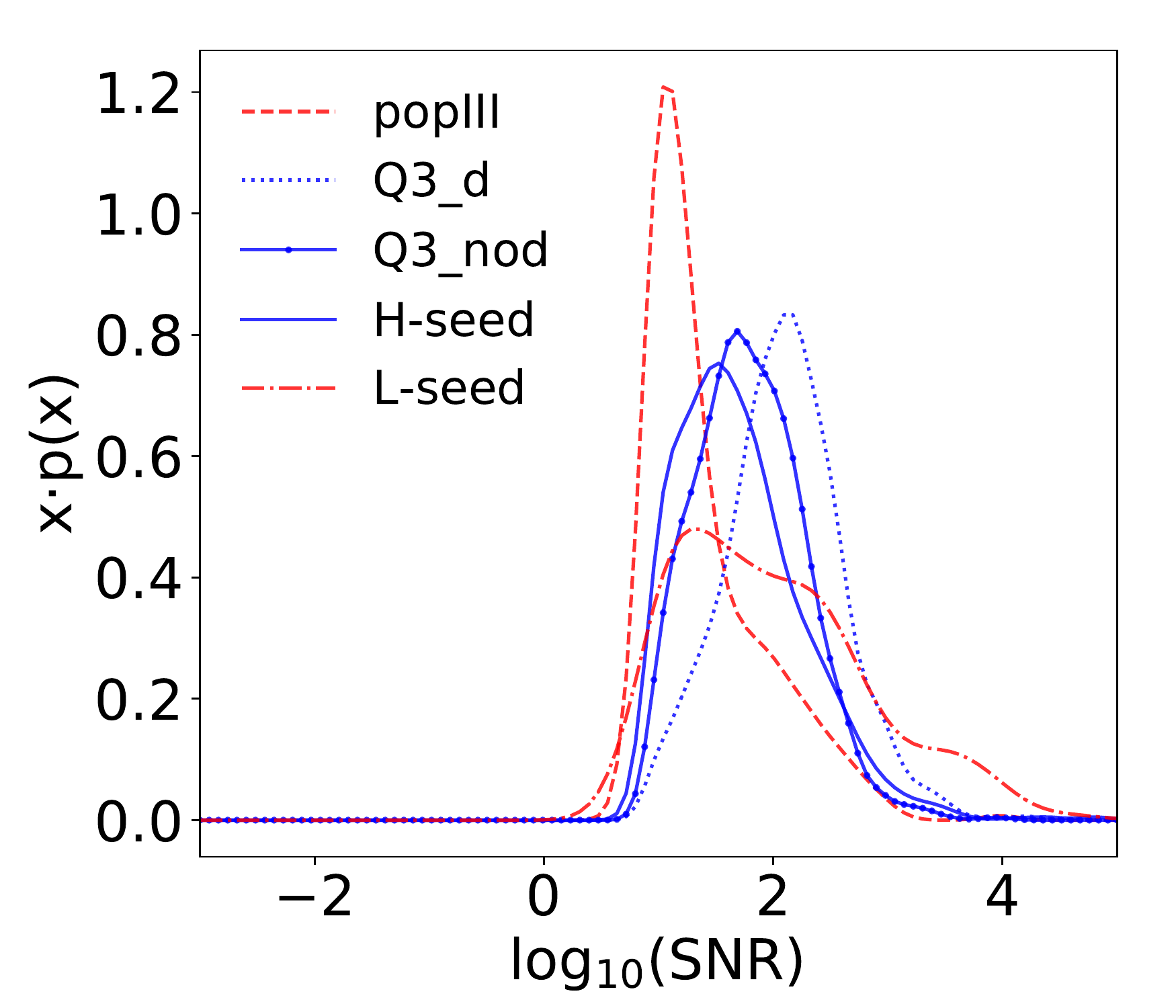}
    \label{snr1_all}
  \end{subfigure}%
  \begin{subfigure}[b]{.30\linewidth}
    \centering
    \includegraphics[width=.99\textwidth,height=5cm]{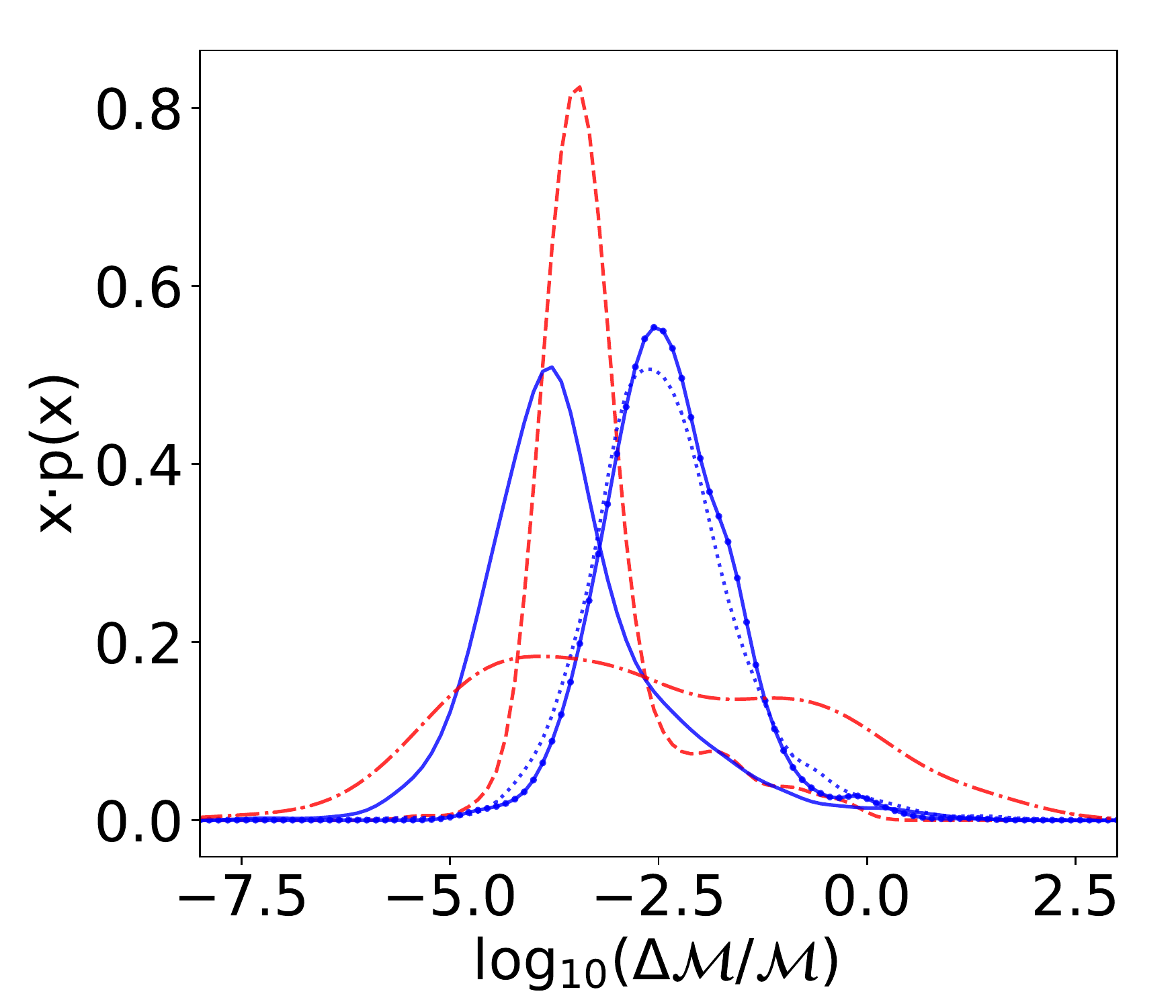}
    \label{cm_all}
  \end{subfigure}%
  \begin{subfigure}[b]{.30\linewidth}
    \centering
    \includegraphics[width=.99\textwidth,height=5cm]{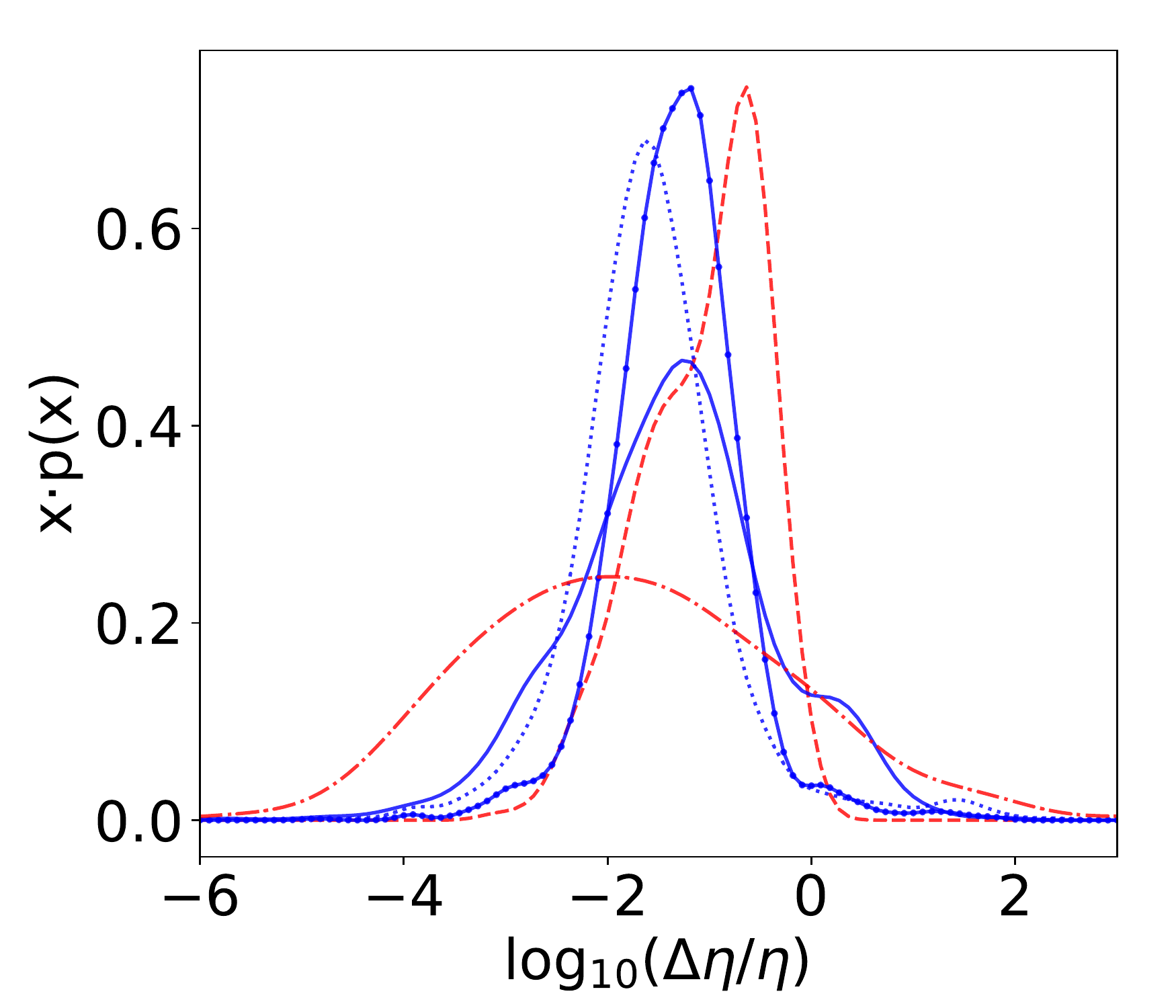}
    \label{eta_all}
  \end{subfigure}\\%
  \begin{subfigure}[b]{.30\linewidth}
    \centering
    \includegraphics[width=.99\textwidth,height=5cm]{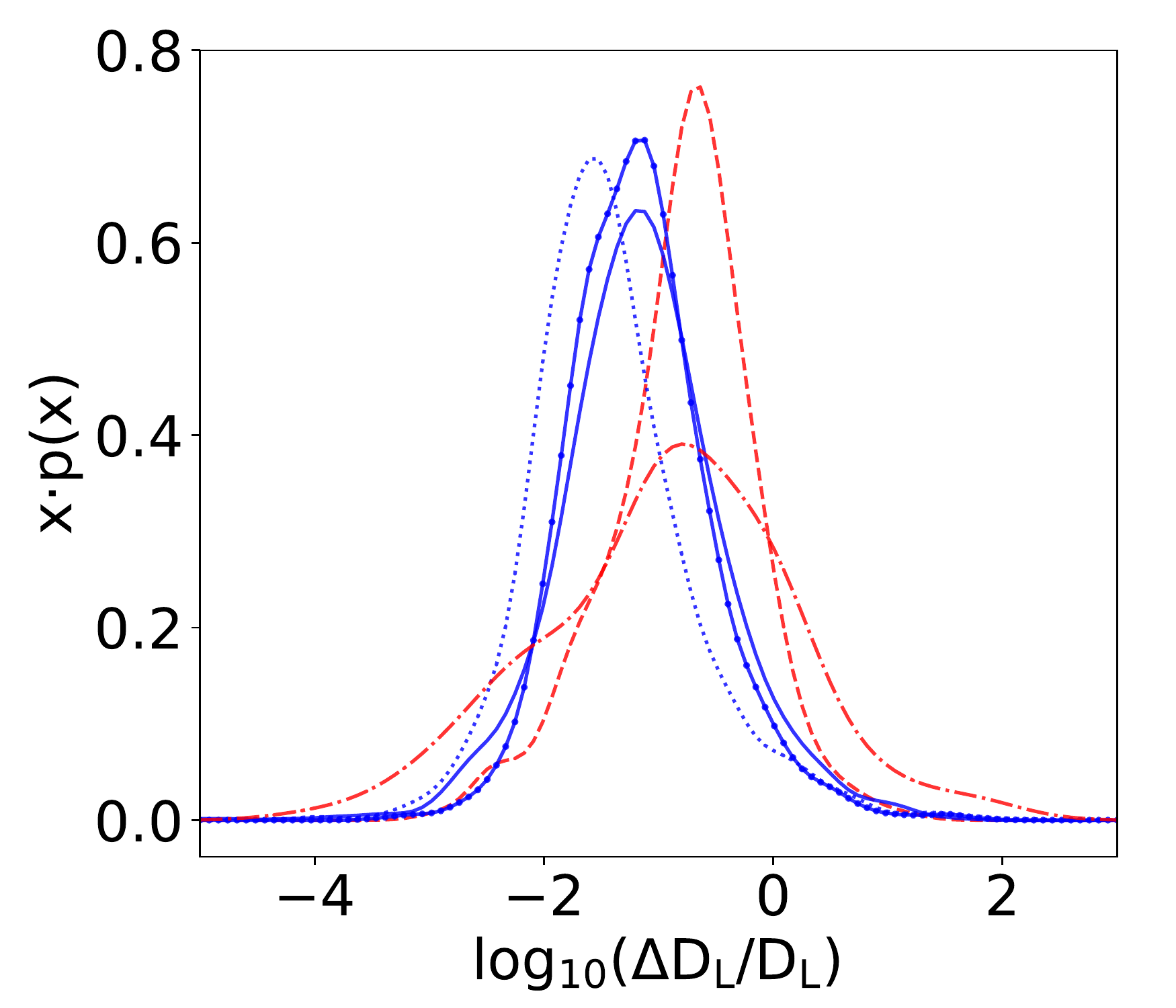}
    \label{dl_all}
  \end{subfigure}%
  \begin{subfigure}[b]{.30\linewidth}
    \centering
    \includegraphics[width=.99\textwidth,height=5cm]{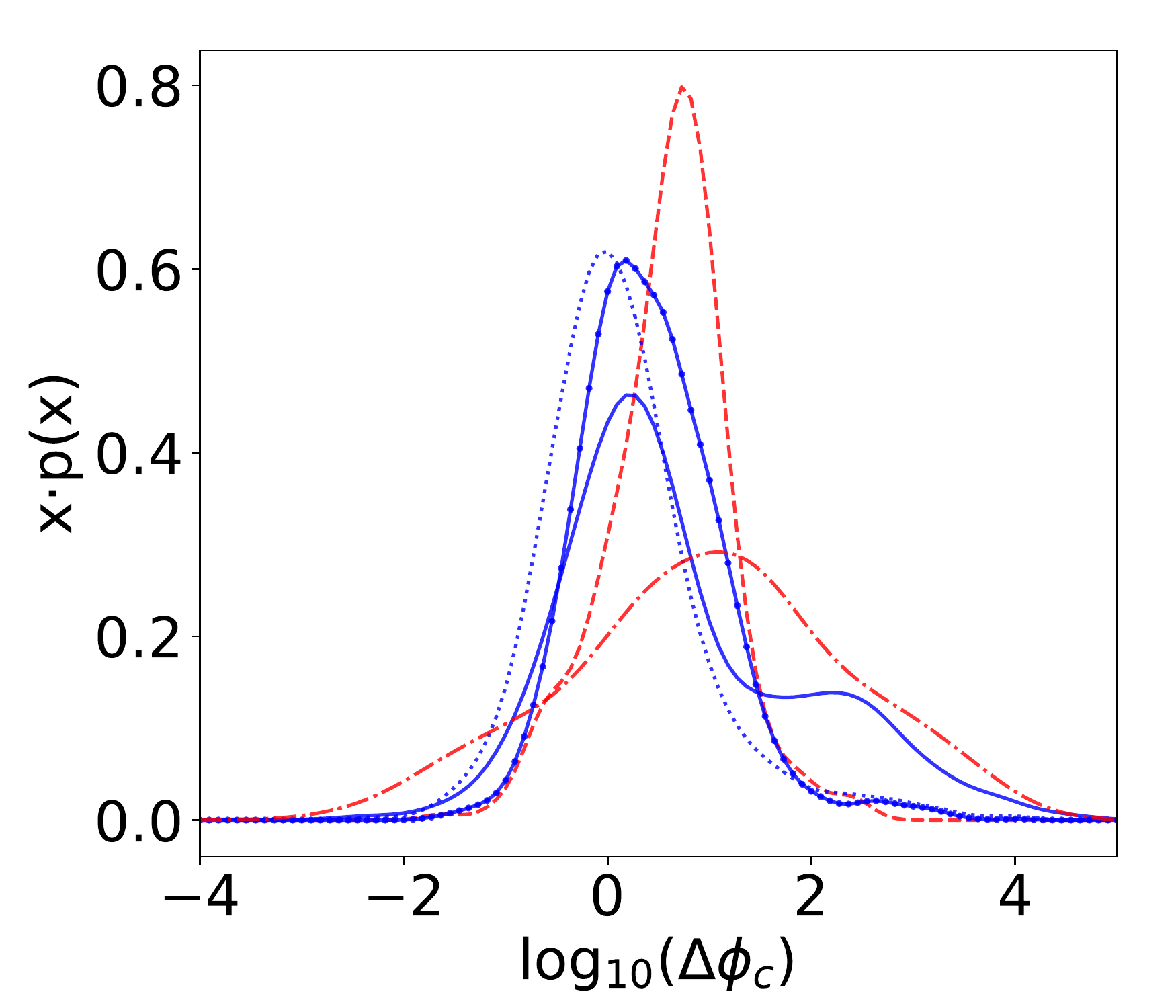}
    \label{phic_all}
  \end{subfigure}%
  \begin{subfigure}[b]{.30\linewidth}
    \centering
    \includegraphics[width=.99\textwidth,height=5cm]{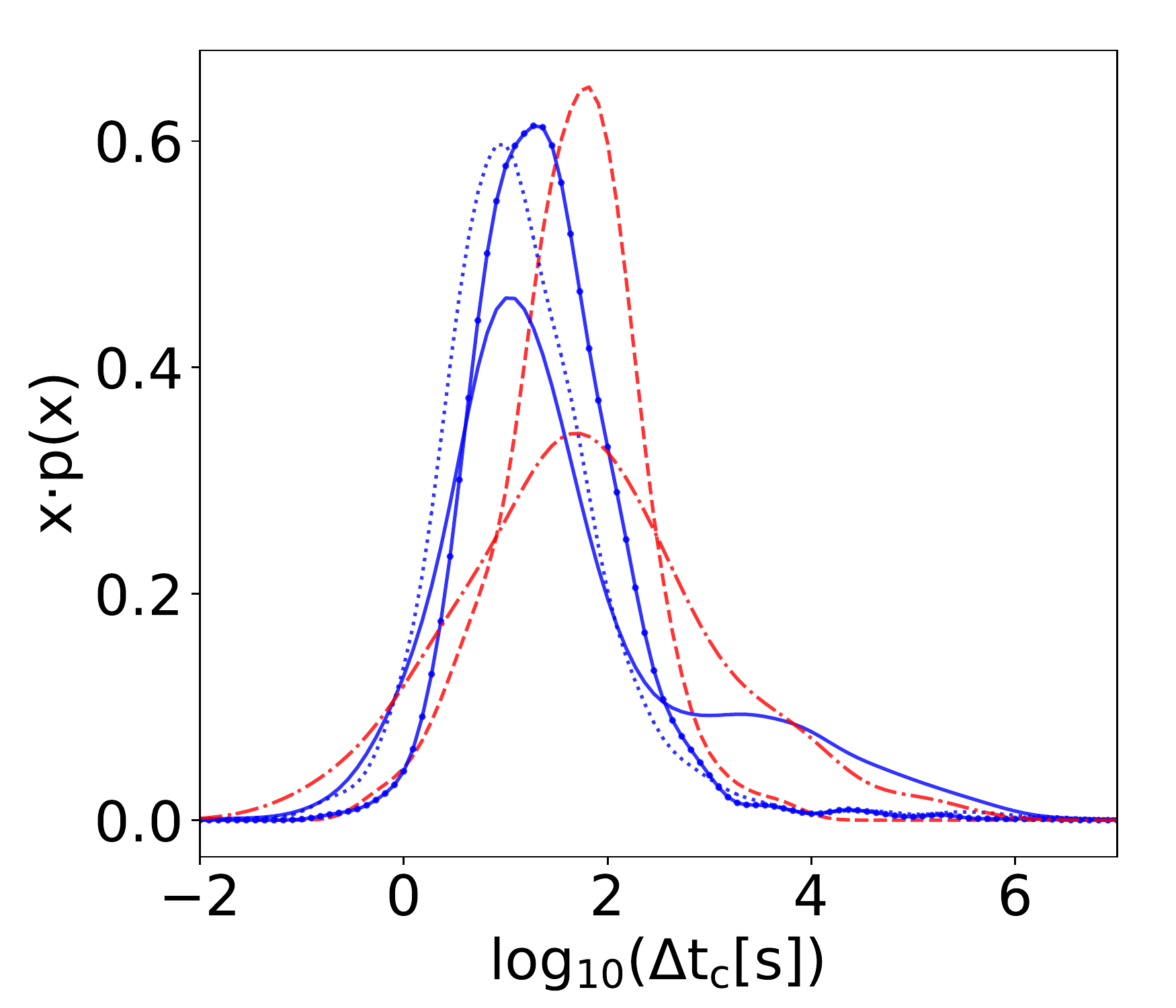}
    \label{tc_all}
  \end{subfigure}\\%
  \begin{subfigure}[b]{.30\linewidth}
    \centering
    \includegraphics[width=.99\textwidth,height=5cm]{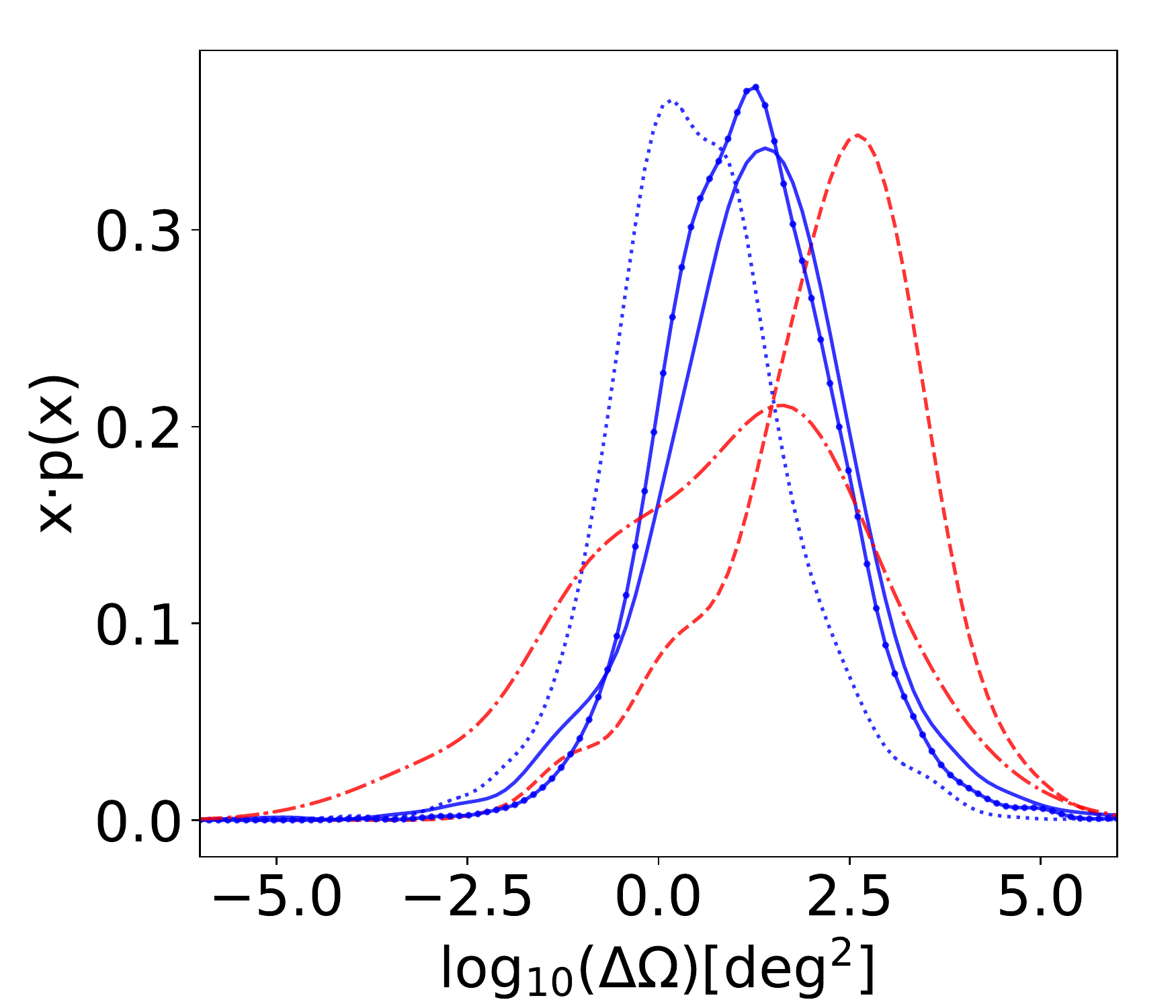}
    \label{loc_all}
  \end{subfigure}%
  \begin{subfigure}[b]{.30\linewidth}
    \centering
    \includegraphics[width=.99\textwidth,height=5cm]{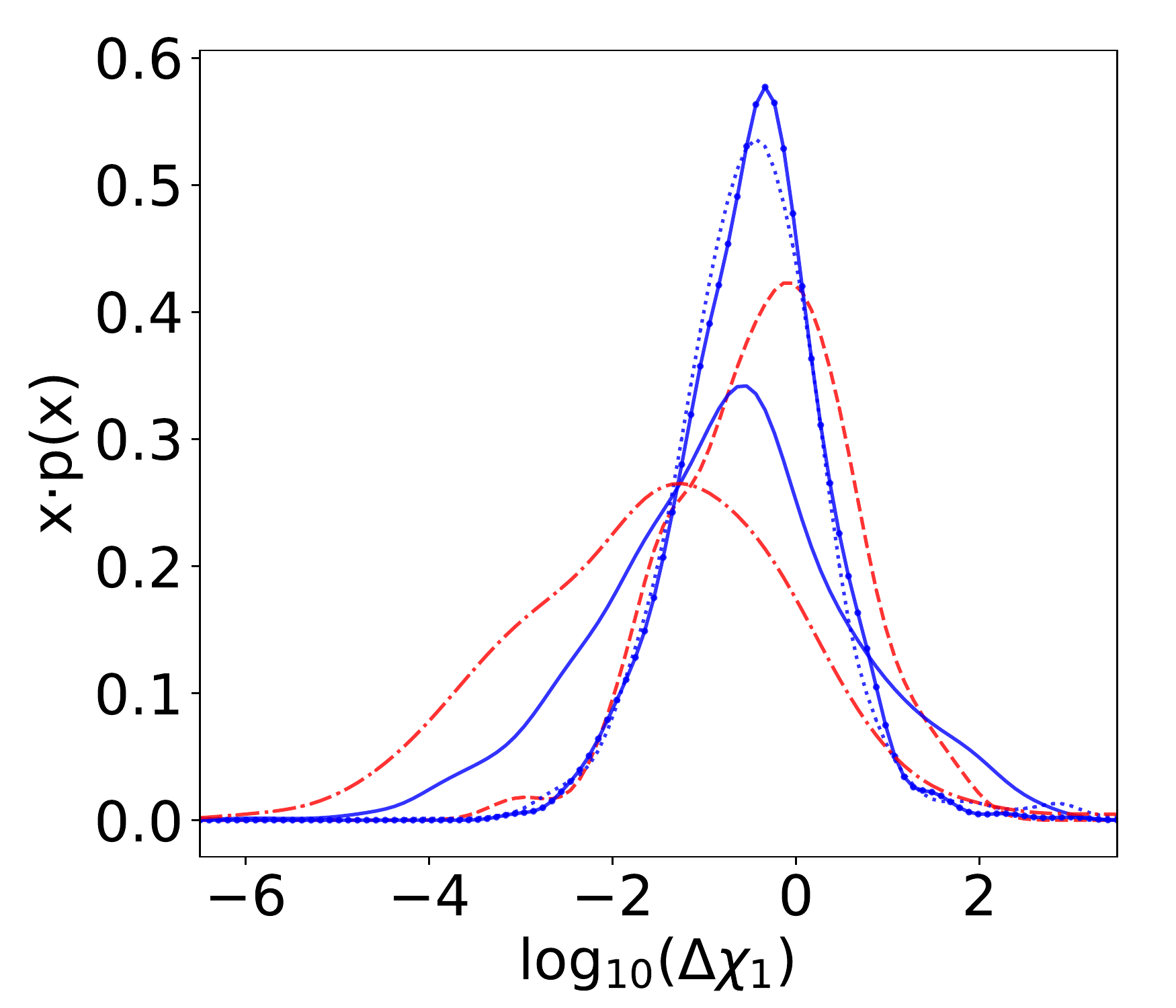}
    \label{chi1_all}
  \end{subfigure}%
  \begin{subfigure}[b]{.30\linewidth}
    \centering
    \includegraphics[width=.99\textwidth,height=5cm]{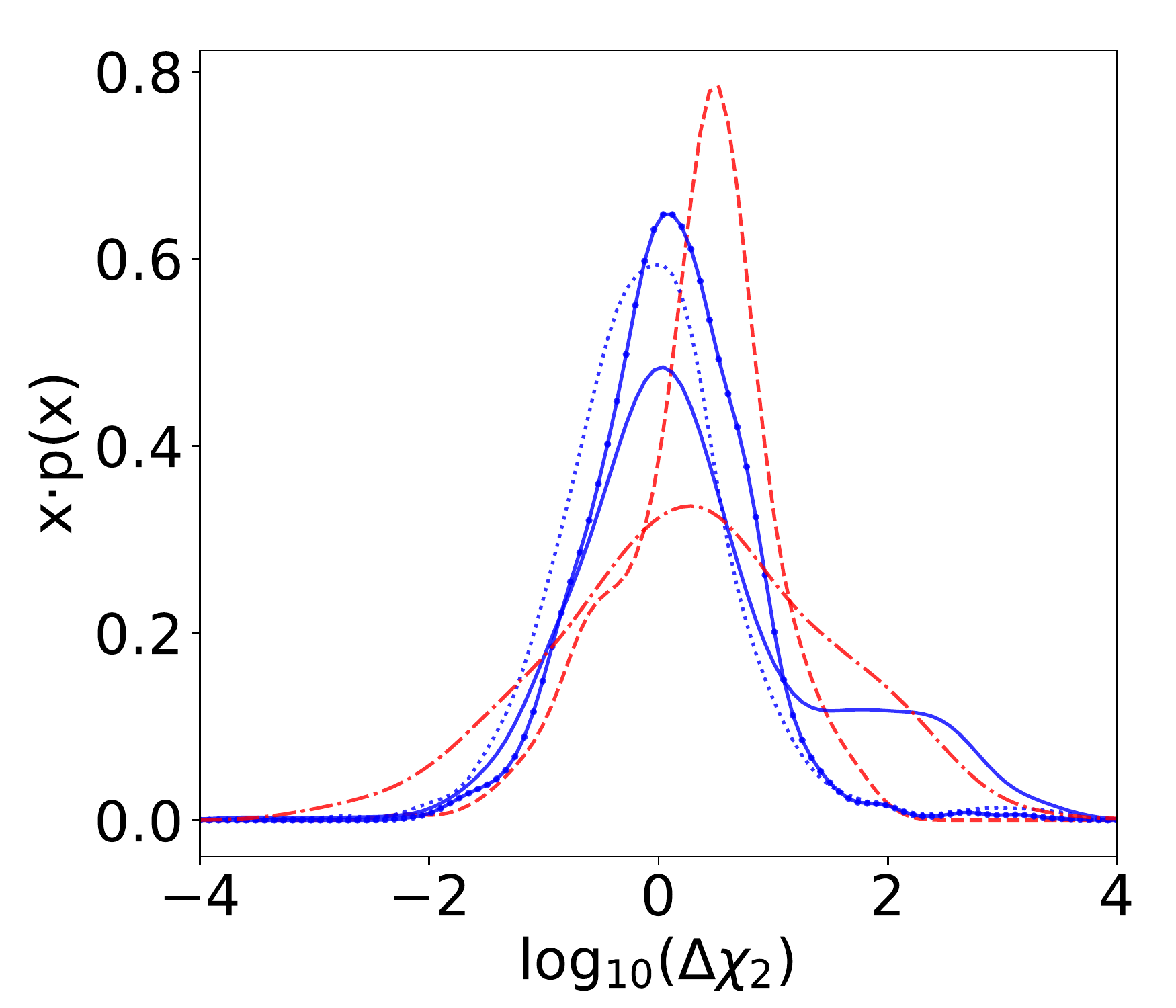}
    \label{chi2_all}
  \end{subfigure}\\%
  \caption{ Probability distribution of SNR and parameter estimation uncertainties for all five models. Source masses and redshifts are taken from mock catalogues of merging \ac{MBH} binaries predicted by the models, whereas other parameters are randomly drawn.}
\label{hist123}
\end{figure}

For detected events from the catalogs of these five models, we perform an analysis on parameter uncertainties, and then we obtain the distribution of uncertainties of all parameters. 
The spin of the source is set to be $\chi_{1,2}\in \text{U}[-1,1]$, the sky location of the source is set to be $\phi\in \text{U}[0,2\pi],\,\cos(\theta)\in \text{U}[-1,1]$; 
the orientation of angular momentum is set to be $\phi_L\in \text{U}[0,2\pi],\,\cos(\theta_L)\in \text{U}[-1,1]$, upon which we can derive the inclination and polarization angles; 
the reference phase at merger is set to be $\phi_c\in \text{U}[0,2\pi]$, and the observation time is set to be $t_c\in \text{U}[0,5]$ year; 
and the chirp mass, symmetric mass-ratio and redshift (luminosity distance) are obtained directly through catalogs derived from models. 
We plot the probability distribution of errors in all the parameters in Fig. \ref{hist123}.
%Again, we apply an \ac{SNR} threshold of 8 first.
We find that TianQin has, in general, a better detection ability for heavy-seed models, compared with the light-seed models. 
This is not surprising, since detected events are typically more massive and at lower redshift in the former (cf. Fig.\ref{A.Klein distributions}). 
This means that the typical detection SNR is higher (cf. the top left panel of Fig. \ref{hist123}), and typical parameter estimation precision scales with the inverse of the SNR. 
Most notably, in the Q3 models, more than 50\% of the detected sources can be located within $\Delta\Omega<10$ deg$^2$ and $\Delta\rm{D_L}/\rm{D_L}<0.03$ (although this does not include weak lensing that can significantly deteriorate the measurement for events at $z>3$ \cite{2005ApJ...629...15H}). 
Conversely, for the light-seed scenarios, those figures are met only for about 20\% of the detections, due to the average lower mass and higher redshift of the sources.

%For the distribution of popIII, Q3\_d and Q3\_nod in Fig.\ref{A.Klein distributions}, we find that the average of the chirp mass in three models are quite different,
%results in different SNR distribution in Fig.\ref{hist123}.
%And the distribution of $\Delta\rm{D_L},\Delta\rm{t_c},\Delta\Omega$ have a strong relation to SNR distribution.

\section{Summary}\label{sec:summary}

We explored the detection and parameter estimation capabilities of TianQin for \ac{MBH} mergers by employing five different models of \ac{MBH} population (namely L-seed, H-seed, popIII, Q3\_d, and Q3\_nod) to characterize optimistic and pessimistic event rates. 
The models feature different techniques (EPS and N-body simulations) for constructing the merger history of dark matter halos and different physical recipes for evolving galaxies and MBHs. 
We find that different models predict vastly different detection scenarios for TianQin, in line with previous investigation focusing on LISA \cite{2007MNRAS.377.1711S,2011PhRvD..83d4036S,Klein16,EAGLE2016}. 
This can be partially attributed to the mass resolution limit in the Millennium-I numerical simulation at the core of the semi-analytic model GABE (the results of which can therefore be taken as lower limits), but it also reflects the large uncertainties in the physics underlying \ac{MBH} formation and evolution, especially at high redshift. 
The detection rate for the current design of TianQin is $\mathcal{O}(1\sim10)$ per year for all the population models except L-seed, while the rate is doubled for twin constellation configuration.
We also showed that if a \ac{MBH} merger with mass $3\times 10^3M_\odot$ happened at the redshift $15$, TianQin could be capable of distinguishing between the heavy- and light-seed models of \ac{MBH}.
Therefore TianQin can shed light on the evolution history of \ac{MBH} population.

TianQin can also trigger an early warning for \ac{MBH} merger by identifying the signal up to 24 hours before merger.
For a merger event that happened at $z=2$, TianQin can generally put constraints on sky localization better than $100 \ {\rm deg}^2$ one day before the merger, falling in the sweet spot of the TianQin sensitivity curve.

To summarize, TianQin is a promising facility to detect \ac{MBH} mergers, has the ability to measure parameters accurately, and has the potential to reveal the nature of the first seed of the \acp{MBH} we see today at the center of galaxies.

\begin{acknowledgments}

This work was supported in part by the National Natural Science Foundation of China (Grants No. 11703098, No. 91636111, No. 11475064, No. 11503007, No. 11690021, and No. 11690022)
and by the European Union's Horizon 2020 research and innovation program under the
Marie Sklodowska-Curie Grant No. 690904. This project has received funding (to E. Barausse) from the European Research Council (ERC) under the European Union's Horizon 2020 research and innovation programme (Grant No. GRAMS-815673; project title “GRavity from Astrophysical to Microscopic Scales”). 
A.S. is supported by the Royal Society. 
We also thanks to Liang Gao, Qi Guo, Vadim Milyukov, Peng-Cheng Li, Gao Jie, and Jian-dong Zhang for insightful comments and discussions. 

\end{acknowledgments}

\bibliographystyle{apsrev4-1}

\bibliography{TQ-MBHB}

\end{document}